%% file: report.tex
\documentclass[10pt, article, letterpaper, onecolumn]{IEEEtran}

\usepackage{amsmath,epsfig,verbatim}
\usepackage{subfigure,algorithm}
\usepackage{graphicx}
\usepackage{amssymb}
\usepackage{url}
\usepackage{color}
\usepackage{mathrsfs}
\usepackage{amsthm,amsfonts}
\usepackage{float}
\usepackage[square, comma, sort&compress, numbers]{natbib}
\usepackage{bm}
\usepackage{algorithmic}
\usepackage{bbm}

\newtheorem{definition}{Definition}
\newtheorem{assumption}{Assumption}
\newtheorem{theorem}{Theorem}
\newtheorem{corollary}{Corollary}
\newtheorem{lemma}{Lemma}

\newcommand*{\QEDA}{\hfill\ensuremath{\blacksquare}}

\begin{document}
\makeatletter
\def\@copyrightspace{\relax}
\makeatother
\title{Paid Peering, Settlement-Free Peering, or Both?}
\author{\IEEEauthorblockN{Xin Wang, Yinlong Xu, Richard T. B. Ma}
\thanks{X. Wang and Y. Xu are with School of Computer Science and Technology, University of Science and Technology of China. R. T. B. Ma is with School of Computing, National University of Singapore. X. Wang's email address is yixinxa@mail.ustc.edu.cn. This work was supported in part by the National Science Foundation of China (No.61379038) and the National Research Foundation, Prime Minister’s Office, Singapore under its Corporate Laboratory@ University Scheme, National University of Singapore, and Singapore Telecommunications Ltd.}
}

\maketitle
\input{sections/Abstract}
\input{sections/Introduction}

\input{sections/Network_Model}

\input{sections/Theory}

\input{sections/Simulation}

\input{sections/RelatedWork}

\input{sections/Conclusions}

\input{sections/Appendix}

\bibliographystyle{abbrv}
\bibliography{paper}

\end{document}

%% file: sections/Abstract.tex
\begin{abstract}
With the rapid growth of congestion-sensitive and data-intensive applications, traditional settlement-free peering agreements with best-effort delivery often do not meet the QoS requirements of content providers (CPs). 
Meanwhile, Internet access providers (IAPs) feel that revenues from end-users are not sufficient to recoup the upgrade costs of network infrastructures. Consequently, some IAPs have begun to offer CPs a new type of peering agreement, called paid peering, under which they provide CPs with better data delivery quality for a fee. 
In this paper, we model a network platform where an IAP makes decisions on the peering types offered to CPs and the prices charged to CPs and end-users.
We study the optimal peering schemes for the IAP, i.e., to offer CPs both the paid and settlement-free peering to choose from or only one of them, as the objective is profit or welfare maximization.
Our results show that 1) the IAP should always offer the paid and settlement-free peering under the profit-optimal and welfare-optimal schemes, respectively, 2) whether to simultaneously offer the other peering type is largely driven by the type of data traffic, e.g., text or video, and 3) regulators might want to encourage the IAP to allocate more network capacity to the settlement-free peering for increasing user welfare.
\end{abstract} 

%% file: sections/Introduction.tex
\section{Introduction}

Internet access providers (IAPs) have constructed massive network platforms to make end-users access the Internet and obtain data from content providers\footnote{We use the term `content provider' in a broad sense that it includes Internet companies, e.g., Facebook \cite{facebook} and Netflix \cite{netflix}, content delivery networks (CDNs), e.g., Akamai \cite{akamai}, and transit ISPs, e.g., Sprint \cite{sprint}.} (CPs). 
Traditionally, data delivery between IAPs and CPs was often based on settlement-free peering agreements \cite{faratin2007complexity}, under which the providers exchange traffic without any form of compensation. As a result, IAPs' revenues were mainly from charges on end-users.
In recent years, however, Internet traffic has been growing more than 50\% per annum \cite{labovitz2010internet} with the rapid popularity of data-intensive services, e.g., online video streaming and cloud-based applications.
To sustain the traffic growth, IAPs need to upgrade their network infrastructures but they feel that the revenues from end-users are often not sufficient to recoup the corresponding costs. 
Meanwhile, such rapid traffic growth has caused serious network congestion, especially during peak hours. Consequently, the best-effort data delivery obtained by the settlement-free peering often does not satisfy the QoS requirements of congestion-sensitive CPs, e.g., Netflix \cite{netflix}.
Owing to these, some IAPs have begun to offer CPs a new type of peering agreement, called paid peering \cite{faratin2007complexity}, under which they provide CPs with a better delivery quality for a fee.
For instance, Comcast \cite{ComcastXFINITY} and Netflix reached a paid peering agreement in 2014 \cite{wyatt2014comcast}, where Comcast offers Netflix a direct connection that requires compensation.
Although the paid peering may increase the revenues of IAPs and improve the data delivery quality of CPs, it has caused concerns over net neutrality \cite{wu2003network}, i.e., whether IAPs should be allowed to charge CPs and differentiate their data traffic.
In 2015, the U.S. Federal Communications Commission (FCC) \cite{FCC} passed the Open Internet Order \cite{FCC_Open} to protect net neutrality. However, existing paid peering agreements were exempt from the ruling. Because the FCC feels that it lacks in-depth background ``in the Internet traffic exchange context''.

Although several prior work \cite{faratin2007complexity,lodhi2014open,ma2017pay} has studied the settlement-free or paid peering, or their impacts on the market participants, e.g., IAPs, CPs, and end-users, some important questions remain unanswered, including the follows.
\begin{itemize}
\item What is the optimal peering scheme for IAPs? More specifically, for maximizing the profits, whether should IAPs offer CPs both of the settlement-free and paid peering to choose from or only one of them?
\item How should regulators make policies on IAPs' peering schemes for protecting the welfare of end-users?
\item How should IAPs and regulators adjust peering schemes and regulatory policies under varying market parameters, e.g., data demand and congestion sensitivity of end-users and capacities of IAPs?

\end{itemize}

There are two challenges in order to address the above questions. 
First, the level of network congestion, which reflects the data delivery quality of a peering type, is an endogenous variable in a network platform and cannot be directly set by an IAP. 
On the one hand, the congestion level is influenced by the data load of network platform. On the other hand, the data usage of end-users and load of network platform are also influenced by the network congestion. 
It is crucial to accurately capture the endogenous congestion so as to faithfully characterize the data delivery quality of a peering type. 
Second, although the peering agreements are between IAPs and CPs, it is not enough to only characterize the IAPs' decisions towards CPs, e.g., pricing on the paid peering and capacity allocation between the settlement-free and paid peering.
We also need to consider the IAPs' charge on end-users because it directly affects the optimal objectives of IAPs and regulators, i.e., the profits and user welfare. Meanwhile, it also affects the data usage of end-users which impacts the level of network congestion and the delivery quality of peering types.

In this work, we model a network platform built by an IAP to answer the above questions. 
We consider that the IAP can offer CPs the settlement-free and paid peering to choose from. 
We measure the quality of a peering type by its congestion level, which is modeled as a function of network capacity and data load.
We consider that the IAP decides the network capacity allocated to the peering types and charges the CPs that use paid peering and end-users for accessing the Internet. 
We capture CPs' choices over the peering types and end-users' data usage under the congestion and pricing parameters.
We derive an endogenous system congestion in an equilibrium.
Based on the equilibrium model, we characterize the optimal strategies, i.e., capacity allocation and pricing decisions, that maximize the IAP's profit or user welfare. We analyze the peering schemes under the optimal strategies. We also evaluate the changes in the optimal strategies under varying system parameters, e.g., data demand and congestion sensitivity of end-users and capacity of the IAP. Our main contributions and results are as follows.

\begin{itemize}
\item We model a network platform in which paid peering, settlement-free peering, or both are offered. We show the existence and uniqueness of a congestion equilibrium (Theorem \ref{the:uniqueness}) and study its changes under varying capacity allocation and pricing parameters (Corollary \ref{cor:pc effect} to \ref{cor:r effect}).
\item We analyze the peering schemes under the profit-optimal strategies (Theorem \ref{the:profit maximization} and Corollary \ref{cor:profit maximization congestion responses} and \ref{cor:profit maximization hazard rate responses}).
We find that to maximize the profits, IAPs need to always offer paid peering. When data traffic is mostly for text (video), they might (might not) want to simultaneously offer settlement-free peering.
\item We analyze the peering schemes under the welfare-optimal strategies (Theorem \ref{the:welfare maximization} and Corollary \ref{cor:welfare maximization congestion responses}). 
We find that to maximize user welfare, settlement-free peering should always be provided. When data traffic is mostly for video (text), paid peering needs (needs not) to be simultaneously provided.
The results suggest that regulators might want to encourage IAPs to allocate more capacity to settlement-free peering. 
\item We observe the changes of the optimal strategies under varying system parameters. We find that with growing user demand and congestion sensitivity, IAPs should allocate more capacity to paid peering and regulators might want to tighten user-side price regulation. However, as IAPs expand their capacities, IAPs and regulators should take the opposite operations.
\end{itemize}

We believe that our model and analysis could help IAPs to choose peering schemes and guide regulators to legislate desirable regulations.

%% file: sections/Network_Model.tex
\section{Network Model}\label{sec:model}

In this section, we model a network platform built by an IAP, which transmits data traffic between CPs and end-users.
In particular, the IAP can offer CPs two types of peering agreements, i.e., settlement-free peering and paid peering.
Since CPs' utilities and preferences on the peering types depend on the data usage of end-users, we first capture the impact of IAP's charge on end-users' population and usage (Section \ref{subsec:model user}).
We then characterize the settlement-free and paid peering types and CPs' choices over them (Section \ref{subsec:model CP}). 
Finally, we describe the IAP's capacity allocation to the peering types and derive a congestion equilibrium of the network platform (Section \ref{subsec:model congestion}).

\subsection{User Population and Data Usage}\label{subsec:model user}

We assume that the IAP adopts usage-based pricing \cite{hande10pricing} to charge end-users and per-unit usage charge is denoted by \(p \in (0,+\infty)\). This pricing scheme is widely used by most wireless IAPs, e.g., AT\&T \cite{att} and T-Mobile \cite{tmobile}, and some wired IAPs, e.g., Comcast \cite{comcastusage}. We consider a continuum of end-users, such that each user is modeled by her average value \(u\) of per-unit data usage. We denote the cumulative distribution function of users' value by \(F_u(\cdot)\) and assume it is continuously differentiable over \(\mathbb{R}_+\)\footnote{In this paper, the sign \(\mathbb{R}_+\) expresses the range \([0,+\infty)\).}. Intuitively, a user would benefit from and subscribe to the Internet access service if and only if her value \(u\) is higher than the price \(p\). Therefore, the population of the active users of the IAP is a function of \(p\), defined by
\begin{equation}\label{eq:user population}
M(p) \triangleq \int_p^{+\infty} dF_u = 1 - F_u(p).
\end{equation}

We consider a continuum of CPs and model each CP by two orthogonal characteristics: its average value \(v\) of per-unit data usage and end-users' average data demand \(w\) on it. We denote the cumulative distribution functions of CPs' value and demand by \(F_v(\cdot)\) and \(F_w(\cdot)\), respectively, and assume they are both continuously differentiable over \(\mathbb{R}_+\).
Data demand of an end-user is her desirable amount of data consumed under a congestion-free network. In fact, when there exists network congestion, e.g., packet delay or drop, end-users' data demand might not be fully filled. We denote the level of network congestion for end-users retrieving CPs' contents by \(\phi\in [0,1]\). We define the end-users' average data usage on a CP of demand \(w\) by \(T(w,\phi) \triangleq wG(\phi)\), i.e., the data demand \(w\) multiplied by a gain factor \(G(\phi)\).

\begin{assumption}\label{ass:gain}
\(G(\phi)\colon [0,1] \mapsto [0,1]\) is a decreasing and continuously differentiable function of \(\phi\).
It satisfies \(G(0) = 1\) and \(G(1) = 0\).
\end{assumption}
Assumption \ref{ass:gain} states that the {\em usage gain} or simply {\em gain} \(G(\phi)\) decreases monotonically when the network congestion \(\phi\) deteriorates. In particular, the gain is one and the end-users' data usage \(T(w,\phi)\) equals the demand \(w\) under no congestion, i.e., \(\phi = 0\).
To characterize the rate of decrease of gain with respect to congestion, {\em elasticity} is often considered.

\begin{definition}
The elasticity of $y$ with respect to $x$, or $x$-elasticity of $y$, is defined by
$\displaystyle\epsilon_x^y \triangleq -\frac{x}{y}\frac{\partial y}{\partial x}$.
\label{def:elasticity}
\end{definition}

Elasticity can be expressed as $\epsilon_x^y = (-{\partial y}/y)/({\partial x}/x)$ and interpreted as the percentage decrease in $y$ (numerator) in response to the percentage increase in $x$ (denominator). In particular, $\epsilon_{\phi}^{G}$ characterizes the percentage decrease in the usage gain in response to the percentage increase in the congestion level. Based on this characterization, different forms of gain functions \(G(\phi)\) can be used to model different congestion sensitivities of data demand. For example, when end-users' data demand is more or less sensitive, i.e., decreases more sharply or gently, with respect to congestion, gain functions with higher or lower elasticities can be adopted, respectively. 

\subsection{Paid Peering and Settlement-Free Peering} \label{subsec:model CP}

We consider that the IAP can offer CPs two types of peering agreements, i.e., paid peering and settlement-free peering, to choose from. The former provides a better data delivery quality than the latter but requires compensation. This peering differentiation scheme has been adopted by many IAPs, e.g., Comcast \cite{ComcastXFINITY} and Verizon \cite{VerizonFios}, in recent years. 
In our model, we use the congestion level of a peering type to reflect its data delivery quality.
We denote the congestion levels of paid and settlement-free peering by \(\phi_h\) and \(\phi_l\), respectively. Because paid peering has better delivery quality than settlement-free peering, we assume the congestion level of paid peering is no higher than that of settlement-free peering, i.e., \(\phi_h\le \phi_l\).
We define the vector of the congestion levels by \(\bm{\phi} = (\phi_h,\phi_l)\). 
We assume the IAP charge the CPs using paid peering a price \(q\in (0,+\infty)\) of per-unit data usage. 

For any CP with value \(v\) and demand \(w\), we define its utility of using paid peering by 
\begin{align*}
\Pi_h(v,w;p,q,\phi_h) \triangleq (v-q)M(p)T(\phi_h,w) = (v-q)M(p)wG(\phi_h),
\end{align*}
i.e., the CP's surplus \((v-q)\) of per-unit data usage multiplied by end-users' aggregate data usage \(M(p)T(\phi_h,w)\) on the CP. Similarly, we define its utility of using settlement-free peering by
\begin{align*}
&\Pi_l(v,w;p,\phi_l) \!\triangleq\! vM(p)T(\phi_l,w) \!=\! vM(p)wG(\phi_l).
\end{align*}
We assume any CP would choose the peering type which induces the higher utility\footnote{Without loss of generality, we assume that a CP would choose paid peering when both peering types induce the same utility.}. Therefore, a CP would choose to use paid peering if and only if 
\begin{equation}\label{eq:gain ratio}
v \ge \underline{v}(q,\bm{\phi}) \triangleq \frac{qG(\phi_h)}{G(\phi_h)-G(\phi_l)}
\end{equation}
where \(\underline{v}(q,\bm{\phi})\) defines the boundary value of CPs choosing different peering types.
Based on the CPs' distributions \(F_v(\cdot)\) and \(F_w(\cdot)\), we define the data loads of paid and settlement-free peering types, i.e., the end-users' aggregate data usages on the CPs that choose them, by
\begin{align}\label{eq:congestion load}
\begin{cases}
\displaystyle D_h(p,q,\bm{\phi}) \triangleq \int_0^{+\infty}\!\!\!\!\int_{\underline{v}(q,\bm{\phi})}^{+\infty}M(p)T(\phi_h,w)dF_vdF_w \vspace{0.05in}\\
\displaystyle D_l(p,q,\bm{\phi}) \triangleq \int_0^{+\infty}\!\!\!\!\int_0^{\underline{v}(q,\bm{\phi})}\!M(p)T(\phi_l,w)dF_vdF_w.
\end{cases}
\end{align}

\subsection{Capacity Allocation and Congestion Equilibrium}\label{subsec:model congestion}
As shown in Equation (\ref{eq:congestion load}), given congestion levels \(\bm{\phi}\), the peering types have induced data loads. 
To accommodate its data load under a certain congestion, each peering type needs to have enough network capacity. We model the capacity needed by a peering type as a function \(C(d,\phi)\) of its data load \(d\) and congestion level \(\phi\). 

\begin{assumption}\label{ass:capacity}
We assume that \(C(d,\phi) = H(\phi)d\), where \(H(\phi)\colon (0,1] \mapsto [0,+\infty)\) is a decreasing and continuously differentiable function of \(\phi\) and satisfies \(H(\phi)\rightarrow +\infty\) as \(\phi \rightarrow 0\).
\end{assumption}

The form of function \(C(d,\phi)\) in Assumption \ref{ass:capacity} captures the capacity sharing \cite{chau2010viability} nature of network services, under which the needed capacity is proportional to the data load, i.e., \(kC(d,\phi) = C(kd,\phi)\). This guarantees CPs will not perceive any difference in terms of congestion level after service partitioning or multiplexing \cite{chau2010viability}. Therefore, the IAP can arbitrarily partition its capacity to multiple service classes (in our case, two peering types) without degradation of congestion level. 
In Assumption \ref{ass:capacity}, we also define a {\em capacity} function \(H(\phi)\) to measure the capacity needed by a peering type that has a unit data load and maintains a congestion level \(\phi\). Intuitively, a peering type needs less capacity as its congestion deteriorates, i.e., \(H(\phi)\) decreases with \(\phi\). 
In particular, when \(H(\phi) = 1/\phi\), the capacity needed by a peering type is \(C(d,\phi) = d/\phi\) which was widely used by prior work \cite{gibbens2000internet,jain2001analysis,richard2014pay,wang2017optimal}.
We denote the inverse function of \(C(d,\phi)\) with respect to \(d\) by \(C^{-1}(c,\phi)\), which can be interpreted as the implied data load under a network capacity \(c\) and observed congestion level \(\phi\). By Assumption \ref{ass:capacity}, \(C^{-1}(c,\phi)\) increases with both \(c\) and \(\phi\). 

We assume the IAP has a total capacity \(c\in (0,+\infty)\) and that it allocates fractions \(r\) and \((1-r)\) of the capacity to the paid and settlement-free peering, respectively, where \(r\in [0,1]\). 
When the IAP makes exogenous price decisions \(p,q\) and capacity allocation decision \(r\), the congestion levels \(\bm{\phi}\) of the peering types can be endogenously determined. We define such a congestion equilibrium of the network platform as follows.

\begin{definition}[Congestion Equilibrium]\label{def:congestion equilibrium}
For any fixed prices \(p,q\), total capacity \(c\) and allocation decision \(r\), the congestion level \(\bm{\phi} = (\phi_h,\phi_l)\) is an equilibrium if and only if
\begin{align}\label{eq:congestion equilibrium}
\begin{cases}
D_h(p,q,\bm{\phi}) = C^{-1}\big(rc,\phi_h\big)\vspace{0.05in}\\
D_l(p,q,\bm{\phi})  = C^{-1}\big((1-r)c,\phi_l\big).
\end{cases}
\end{align} 
\end{definition}

In Equation (\ref{eq:congestion equilibrium}) of Definition \ref{def:congestion equilibrium}, the left-hand sides are the induced data loads of the peering types given the price decisions \(p,q\) and congestion levels \(\bm{\phi}\), and the right-hand sides are their implied data loads under the capacity \(c\) and allocation decision \(r\). Under an equilibrium, both equal the end-users' aggregate data usages on the CPs using the peering types.

\begin{theorem}\label{the:uniqueness}
Under Assumption \ref{ass:gain} and \ref{ass:capacity}, for any fixed price decisions \(p,q\), total capacity \(c\) and allocation decision \(r\), there always exists a unique equilibrium for the network platform.
\end{theorem} 

Theorem \ref{the:uniqueness} states that under minor assumptions of the usage gain (Assumption \ref{ass:gain}) and the required capacity (Assumption \ref{ass:capacity}), the existence and uniqueness of equilibrium can be guaranteed. Based on Theorem \ref{the:uniqueness}, we denote the unique equilibrium congestion by \(\bm{\varphi} = (\varphi_h,\varphi_l)\). 

We denote the IAP's {\em strategy} by \(\theta \triangleq (p,q,r)\), i.e., the triple of the price and allocation decisions. Because the equilibrium congestion is determined by the strategy \(\theta\) and the total capacity \(c\), we also denote it by \(\bm{\varphi}(\theta,c)\). We denote the corresponding data loads of the paid and settlement-free peering types, respectively, by 
\[d_h(\theta,c) \!\triangleq D_h\big(p,q,\bm{\varphi}(\theta,c)\big) \ \text{and} \ \ d_l(\theta,c) \!\triangleq D_l\big(p,q,\bm{\varphi}(\theta,c)\big).\]
We define the total data load of the network platform by \(d_t(\theta,c) \triangleq d_h(\theta,c) + d_l(\theta,c)\), i.e., the summation of the data loads of the two peering types.
The following three corollaries show how the equilibrium congestion and data loads change when the strategy \(\theta\) changes.

\begin{corollary}\label{cor:pc effect}
For any fixed price \(q\in (0,+\infty)\) and allocation decision \(r\in (0,1)\), the congestion levels \(\varphi_h,\varphi_l\) and data loads \(d_h,d_l\) all decrease with the user-side price \(p\).
\end{corollary}

Corollary \ref{cor:pc effect} tells that the congestion levels and data loads of the peering types will decrease if the user-side price increases. Intuitively, when the IAP raises the user-side price, the population of the active end-users will decline, which further reduces the congestion levels under equilibrium and corresponding data loads of the peering types. 

\begin{corollary}\label{cor:q effect}
For any fixed price \(p\in (0,+\infty)\) and allocation decision \(r\in (0,1)\), the congestion level \(\varphi_h\) (\(\varphi_l\)) and data load \(d_h\) (\(d_l\)) both decrease (increase) with the price \(q\) of paid peering.
\end{corollary}

Corollary \ref{cor:q effect} tells that the congestion level and data load of paid (settlement-free) peering type decrease (increase) if the price \(q\) of paid peering increases. Because some CPs that choose paid peering will shift to settlement-free peering under a higher price \(q\). This result also tells that the IAP could use the price of paid peering as a means to adjust the service segment of low or high congestion, i.e., the set of CPs that choose paid or settlement-free peering, respectively.

\begin{corollary}\label{cor:r effect}
For any fixed prices \(p,q\in (0,+\infty)\), the congestion level \(\varphi_l\) and data load \(d_h\) increase with the allocation decision \(r\) and the data load \(d_l\) decreases with the allocation decision \(r\). Besides, it satisfies 1) \(\varphi_l = \varphi_h\) if and only if \(r=0\) and 2) \(\varphi_l = 1\) if and only if \(r=1\).
\end{corollary}

Corollary \ref{cor:r effect} states that the congestion level \(\varphi_l\) of settlement-free peering type will increase if the capacity fraction \(r\) allocated to paid peering increases. In particular, when the IAP allocates all its capacity to the settlement-free peering type, i.e., \(r=0\), the congestion levels \(\varphi_l, \varphi_h\) of the two peering types are equal, under which no CPs will use paid peering and the price \(q\) does not play a role. 
In such a case, we say the IAP's peering scheme is {\em pure free peering}, i.e., to only offer the settlement-free peering type. 
At the other extreme, when the IAP allocates all its capacity to the paid peering type, i.e., \(r=1\), the congestion level \(\varphi_l\) of the settlement-free peering type is one, which can be interpreted as a termination of the settlement-free peering type. 
In such a case, we say the IAP's peering scheme is {\em pure paid peering}, i.e., to only offer the paid peering type. 
Otherwise, when the paid and settlement-free peering types are both allocated with positive capacity, i.e., \(r\in (0,1)\), they both include active CPs. 
In such a case, we say the IAP's peering scheme is {\em hybrid peering}, i.e., to simultaneously offer the paid and settlement-free peering types.

%% file: sections/Theory.tex
\section{Peering Schemes under Optimal Strategies}
In the previous section, we constructed a network model under which the IAP can offer CPs differentiated peering types.
Based on this model, we explore the optimal strategies that maximize the IAP's profit or end-users' welfare in this section. We will show that the peering schemes under the optimal strategies (pure paid, pure free, or hybrid peering) largely depend on the type of data traffic, e.g., text or video, and the characteristic of network capacity.
In particular, we first show the embodiments of traffic type and capacity characteristic in our model as a preliminary (Section \ref{subsec:type and characteristic}).
We then study the IAP's profit-optimal strategy and its corresponding peering scheme under various network conditions (Section \ref{subsec:profit-optimal pricing}).
Finally, we study the welfare-optimal strategy and compare it with the profit-optimal counterpart to derive regulatory implications (Section \ref{subsec:welfare-optimal pricing}).

\subsection{Traffic Type and Capacity Characteristic}\label{subsec:type and characteristic}

As the demanders of network service, end-users' data usage is discounted under network congestion. In Section \ref{subsec:model user}, we captured the discount of usage by the gain function \(G(\phi)\) and the congestion sensitivity of usage by the elasticity of gain \(\epsilon^G_\phi\). Furthermore, when the end-users' usage is for different types of data traffic, e.g., text or video, the congestion elasticity of gain may have very different properties. 
In particular, text traffic such as file sharing is usually insensitive to mild congestion but cannot tolerate severe congestion. Conversely, video traffic such as online video streaming is quite sensitive to mild congestion. Based on these characteristics, we assume that the usage gain \(G(\phi)\) of text (video) traffic decreases concavely (convexly) with the congestion level \(\phi\) and its elasticity \(\epsilon^G_\phi\) increases (decreases) with the congestion level \(\phi\). Under such assumptions, the type of data traffic can be embodied in the monotonicity of the elasticity of gain.

As the supplier of network service, the IAP is required to provide enough capacity to guarantee the service quality. In Section \ref{subsec:model CP}, we captured the capacity requirement by the capacity function \(H(\phi)\). Because it always requires more capacity to maintain lower congestion level, we assumed that \(H(\phi)\) decreases with the congestion level \(\phi\). Besides the monotonicity of the capacity function, characteristics of network capacity are also reflected in the monotonicity of the congestion elasticity \(\epsilon^H_\phi\), which depends on the network technology adopted by the IAP. An increasing (decreasing) elasticity function \(\epsilon^H_\phi\) of the congestion level \(\phi\) means that as \(\phi\) is higher, the required capacity is more (less) sensitive to the congestion.

In summary, the gain function \(G(\phi)\) and the capacity function \(H(\phi)\) characterize the impact of network congestion on the end-users and the IAP, respectively.
With the development of CPs' content services and IAPs' network technologies, end-users' data traffic type and IAPs' capacity characteristic are continuously changing. These changes can be reflected in the monotonicity of the congestion elasticities of the gain and capacity functions which affect the peering schemes of the optimal strategies as shown in the following two subsections.

\subsection{Profit-optimal Strategy}\label{subsec:profit-optimal pricing}
In this subsection, we study the optimal strategy and the corresponding peering scheme which maximize the IAP's profit. We consider that the IAP incurs a cost of \(k \in (0,+\infty)\) by serving per-unit data traffic, which models the recurring maintenance and utility cost like electricity. We define the IAP's profit by \(U(\theta,c,k) \triangleq (p+q-k) d_h(\theta,c) + (p-k) d_l(\theta,c)\),
where the first and second terms are its profits from the data traffic of paid and settlement-free peering, respectively. Under any given total capacity \(c\) and per-unit traffic cost \(k\), the IAP can maximize its profit \(U\) by determining the strategy \(\theta\), i.e., the triple \((p,q,r)\), that solves the optimization problem:
\begin{align*}
\underset{\theta}{\text{maximize}} \quad & U(\theta,c,k) = (p\!+\!q\!-\!k) d_h(\theta,c) + (p\!-\!k) d_l(\theta,c) \\
\text{subject to} \quad & \theta\in (0,+\infty) \!\times\! (0,+\infty) \!\times\! [0,1].
\end{align*}

By solving the above optimization problem, we characterize the IAP's profit-optimal strategy as follows.

\begin{theorem}\label{the:profit maximization}
If a strategy \(\theta = (p,q,r)\) maximizes the IAP's profit \(U\), then its capacity allocation decision must not be zero, i.e., \(r\neq 0\), and the following conditions must hold:\\
1) the IAP's revenue from end-users satisfies that
\begin{equation}\label{eq:profit maximization average}
\displaystyle p d_t = (p+q-k)d_h\epsilon^{d_h}_p + (p - k) d_l \epsilon^{d_l}_p. 
\end{equation}
2) the ratio of the IAP's profits from the paid and settlement-free peering types satisfies that
\begin{equation}\label{eq:profit maximization ratio}
\frac{(p+q-k)d_h}{(p-k)d_l} \!=\! \frac{-pd_t\epsilon^{d_l}_q+qd_h\epsilon^{d_l}_p}{pd_t\epsilon^{d_h}_q-qd_h\epsilon^{d_h}_p} \!\!
\begin{cases}
\!=\! -\displaystyle\frac{\epsilon^{d_l}_r}{\epsilon^{d_h}_r} \  \text{if} \  r\in (0,1);\vspace{0.05in}\\
\!\ge\! -\displaystyle\frac{\epsilon^{d_l}_r}{\epsilon^{d_h}_r}\  \text{if} \   r\! =\! 1.
\end{cases}
\end{equation}
\end{theorem} 

Theorem \ref{the:profit maximization} states that the capacity allocation decision of a profit-maximizing IAP must not be zero, i.e., the peering scheme under any profit-optimal strategy must not be the pure free peering. 
In fact, the IAP could always be better off by switching from the pure free peering scheme to the hybrid peering scheme.
Intuitively, when the IAP adopts the pure free peering scheme, i.e., \(r=0\), its revenue is all from end-users. 
However, when the IAP switches to the hybrid peering scheme, i.e., \(r\in (0,1)\), it simultaneously offers the paid and settlement-free peering types and some high-value CPs would shift from the settlement-free peering type to the paid peering type. As a result, the IAP could extract additional revenue from the high-value CPs and thus obtain higher total profit.
Notice that from the perspective of profit maximization, although the hybrid peering scheme is always better than the pure free peering scheme, this kind of relationship does not exist between the pure paid and pure free peering schemes. 
When the IAP shifts from the pure free peering scheme to the pure paid peering scheme, i.e., \(r=1\), some low-value CPs which cannot afford the paid peering type have to exit the network platform. Consequently, the IAP can no longer charge end-users for their data usages on the low-value CPs, although it can generate additional revenue from the high-value CPs using the paid peering type. As a result, the comprehensive effect of these changes may either increase or decrease the IAP's total profit.

Theorem \ref{the:profit maximization} also gives the necessary conditions, i.e., Equation (\ref{eq:profit maximization average}) and (\ref{eq:profit maximization ratio}), that the profit-optimal strategies need to meet. 
Equation (\ref{eq:profit maximization average}) shows the relationship among the IAP's revenue from end-users (left-hand side), its profits from the data traffic of paid and settlement-free peering types, and the elasticities of the data loads with respect to the user-side price.
Equation (\ref{eq:profit maximization ratio}) characterizes the relationship among the ratio of the profits from the data traffic of paid and settlement-free peering types and the elasticities of the data loads with respect to the price and allocation decisions.

By far we have shown that the peering scheme of a profit-maximizing IAP must not be the pure free peering. Furthermore, Corollary \ref{cor:profit maximization congestion responses} and \ref{cor:profit maximization hazard rate responses} tell when the scheme can only be the pure paid peering and the hybrid peering, i.e., \(r=1\) and \(r\in (0,1)\), respectively. 

\begin{corollary}\label{cor:profit maximization congestion responses}
For any profit-optimal strategy, its capacity allocation decision must be one, if \(H(\epsilon^H_\phi/\epsilon^G_\phi+1)\) is an increasing function of the congestion level \(\phi\).
\end{corollary} 

Corollary \ref{cor:profit maximization congestion responses} provides a sufficient condition for the peering scheme under any profit-optimal strategy to be the pure paid peering: the function \(H(\epsilon^H_\phi/\epsilon^G_\phi+1)\) is increasing in the congestion level \(\phi\). In this condition, \(\epsilon^H_\phi/\epsilon^G_\phi\) can be interpreted as a metric of {\em relative gain elasticity of capacity}, i.e., the ratio of the percentage increases in the required capacity and usage gain in response to the percentage decrease in the congestion level. Because the gain function \(G\) and the capacity function \(H\) both decrease with the congestion level \(\phi\), if \(H(\epsilon^H_\phi/\epsilon^G_\phi+1)\) is an increasing function of \(\phi\), the relative gain elasticity of capacity \(\epsilon^H_\phi/\epsilon^G_\phi\) must increase with \(\phi\). This means that as the congestion level is lower, the required capacity is less elastic to the usage gain, or equivalently, the usage gain is more elastic to the required capacity. 
Under such a condition, when the IAP increases the capacity allocated to the paid peering type that maintains a lower congestion level, the increase of the profit from the data traffic of paid peering type is larger than the decrease of the profit from the data traffic of settlement-free peering type. As a result, the IAP's total profit would increase. Thus, the IAP should only offer the paid peering type for optimizing its total profit.

Furthermore, we see that whether the sufficient condition in Corollary \ref{cor:profit maximization congestion responses} holds largely depends on the monotonicity of the elasticity of capacity \(\epsilon^H_\phi\) and the monotonicity of the elasticity of gain \(\epsilon^G_\phi\). The former and the latter reflect the characteristic of system capacity and the type of data traffic, respectively, as we assumed in Section \ref{subsec:type and characteristic}.
When \(\epsilon^H_\phi\) decreases with congestion and \(\epsilon^G_\phi\) increases with congestion, e.g., data traffic is mainly for text content, the sufficient condition must not hold. Under such a case, the profit-optimal peering schemes are often the hybrid peering, which will be shown in Section \ref{sec:simulation}.
Conversely, when \(\epsilon^H_\phi\) increases with congestion or \(\epsilon^G_\phi\) decreases with congestion, e.g., data traffic is mostly for online video, the sufficient condition may hold, under which the profit-optimal peering schemes are the pure paid peering. 

\begin{definition}
The hazard rate of a cumulative distribution function \(F(x)\) is defined by \(\tilde{F}(x) = F'(x)/\left[1-F(x)\right]\). 
\end{definition}

The hazard rate captures the proportion of the complementary cumulative distribution \(1-F(x)\) that is reduced due to a marginal increase of \(x\) and measures the rate of decrease in \(1-F(x)\) at the value \(x\). 
In particular, for the cumulative distributions of users' value \(F_u(x)\) and CPs' value \(F_v(y)\), the hazard rates \(\tilde{F}_u(x)\) and \(\tilde{F}_v(y)\) measure the rates of decrease in the population of users and CPs whose values are higher than \(x\) and \(y\), respectively.

\begin{corollary}\label{cor:profit maximization hazard rate responses}
For any profit-optimal strategy, its capacity allocation decision must be in the interval \((0,1)\), if the inequality \(\tilde{F}_u(p) < \tilde{F}_v(q)\) holds for any prices \(p,q\in (0,+\infty)\).
\end{corollary} 

Corollary \ref{cor:profit maximization hazard rate responses} provides a sufficient condition for the peering scheme under any profit-optimal strategy to be the hybrid peering: the hazard rate of the distribution of users' value \(\tilde{F}_u(p)\) is lower than that of CPs' value \(\tilde{F}_v(q)\) for any prices \(p\) and \(q\). This condition means that with the increases in any two-sided prices \(p\) and \(q\), the rate of decrease of the population of the users whose values are higher than \(p\) is always slower than that of the CPs whose values are higher than \(q\). Under this condition, the user side has a higher value on data usages than the CP side. When the IAP adopts the pure paid peering scheme, some low-value CPs will quit the IAP's network platform and the IAP will lose lots of charges from the high-value users for their data usages on the low-value CPs. As a result, the IAP's profit cannot be maximized. By contrast, when the IAP adopts the hybrid peering scheme, all CPs will connect to the IAP by the paid or settlement-free peering type and the IAP can fully charge the high-value users, under which the IAP's profit can be maximized.

\vspace{0.03in}
\textbf{Summary of Implications}: The theoretical results in this subsection could guide IAPs in choosing peering schemes. 
In particular, IAPs should always provide the paid peering type for optimizing their profits (by Theorem \ref{the:profit maximization}), because it can extract additional revenue by charging the high-value CPs. 
Furthermore, when the CP side is more sensitive to price than the user side or network traffic is mainly for text content, the peering schemes under the profit-optimal strategies are usually to simultaneously offer the paid and settlement-free peering types (by Corollary \ref{cor:profit maximization hazard rate responses}). 
However, as online video streaming increases constantly and users become more sensitive to network congestion, it may change to be to only offer the paid peering type (by Corollary \ref{cor:profit maximization congestion responses}). 

\subsection{Welfare-optimal Strategy}\label{subsec:welfare-optimal pricing}

In this subsection, we explore the welfare-optimal strategy and the corresponding peering scheme. We also contrast them with the profit-optimal counterparts so as to draw implications on desirable regulations from a regulatory perspective. 

For any user-side price \(p\), a user of value \(u\) gets the surplus \((u-p)\) for one unit data usage, and therefore, the total surplus of all users, when each of them generates one unit data usage, can be defined by
\[S(p) \triangleq \int_p^{+\infty} (u-p) dF_u\]
where \(F_u(\cdot)\) is the cumulative distribution function of users' value. Furthermore, the per-user average surplus for one unit data usage can be defined by \(s(p) \triangleq S(p)/M(p)\) where \(M(p)\) is the user population defined by Equation (\ref{eq:user population}). Accordingly, we define the total user welfare by 
\begin{equation}\label{eq:user welfare}
W(\theta,c) \triangleq s(p)d_t(\theta,c) = s(p)\big[d_h(\theta,c)+d_l(\theta,c)\big],
\end{equation}
i.e., the users' average per-unit usage surplus \(s(p)\) multiplied by the total data usages \(d_t(\theta,c)\).
Under any given capacity \(c\), we can maximize the user welfare \(W\) by determining the strategy \(\theta\), i.e., the triple \((p,q,r)\), that solves the optimization problem:
\begin{align*}
\underset{\theta}{\text{maximize}} \quad & W(\theta,c) = s(p) d_t(\theta,c) \\
\text{subject to} \quad & \theta\in (0,+\infty) \!\times\! (0,+\infty) \!\times\! [0,1].
\end{align*}

By solving the above optimization problem, we characterize the welfare-optimal strategy as follows.

\begin{theorem}\label{the:welfare maximization}
If a strategy \(\theta = (p,q,r)\) maximizes the user welfare \(W\), then its capacity allocation decision must not be one, i.e., \(r\neq 1\), and the elasticities of the total data load \(d_t\) with respect to the decisions \(p,q,\) and \(r\) must satisfy:
\begin{equation}\label{eq:welfare maximization}
\epsilon^{d_t}_p  = - \epsilon^s_p,\ \epsilon^{d_t}_q = 0,\ \text{and}\ \, \epsilon^{d_t}_r
\begin{cases}
= 0 & \text{if} \ \, r\in (0,1);\\
\ge 0 & \text{if} \ \, r = 0.
\end{cases}
\end{equation}
\end{theorem} 

Theorem \ref{the:welfare maximization} shows that the capacity allocation decision of a welfare-maximizing IAP must not be one, i.e., the peering scheme under any welfare-optimal strategy must not be the pure paid peering.
By Equation (\ref{eq:user welfare}), under any given user-side price, the IAP's capacity allocation should maximize its total data load for optimizing user welfare. As mentioned before, however, when the IAP adopts the pure paid peering scheme (i.e., only offers the paid peering type), it will lose the data usages on the low-value CPs which cannot afford the paid peering type and thus the total data load cannot be maximized.
This result implies that to protect user welfare, the settlement-free peering type should be retained for guaranteeing that all CPs connect to the IAP. In other words, the IAP should be encouraged to employ the pure free or hybrid peering scheme. 
 
Besides, Equation (\ref{eq:welfare maximization}) in Theorem \ref{the:welfare maximization} gives necessary conditions for strategies to be welfare-optimal by characterizing the elasticities \(\epsilon^{d_t}_p, \epsilon^{d_t}_q\) and \(\epsilon^{d_t}_r\) of the total data load with respect to the strategy parameters. In particular, the user-side price elasticity \(\epsilon^{d_t}_p\) is opposite to the elasticity \(\epsilon^s_p\) of the user average surplus and the CP-side price elasticity \(\epsilon^{d_t}_q\) is zero. The capacity allocation elasticity \(\epsilon^{d_t}_r\) is zero or non-negative, when the peering scheme is the hybrid or pure free peering, respectively.

Theorem \ref{the:welfare maximization} has implied that the peering scheme of a welfare-maximizing IAP can only be the pure free or hybrid peering. Furthermore, Corollary \ref{cor:welfare maximization congestion responses} tells when the scheme must be the pure free peering, i.e., \(r=0\). 

\begin{corollary}\label{cor:welfare maximization congestion responses}
For any welfare-optimal strategy, its capacity allocation decision must be zero, if \(H(\epsilon^H_\phi/\epsilon^G_\phi+1)\) is a decreasing function of the congestion level \(\phi\).
\end{corollary} 

Corollary \ref{cor:welfare maximization congestion responses} provides a sufficient condition for the peering scheme of any welfare-optimal strategy to be the pure free peering: the function \(H(\epsilon^H_\phi/\epsilon^G_\phi+1)\) is decreasing in the congestion level \(\phi\). Specifically, if the relative gain elasticity of capacity \(\epsilon^H_\phi/\epsilon^G_\phi\) decreases with the congestion level \(\phi\), the sufficient condition must hold because the capacity function \(H\) decreases with \(\phi\).
Under such a condition, when the IAP allocates more capacity to the paid peering type, the increase of the data traffic of paid peering type is smaller than the decrease of the data traffic of settlement-free peering type. As a result, without changing the user-side price, the IAP's total data load will decrease and thus user welfare will decrease. Therefore, for optimizing user welfare, the peering scheme must be the pure free peering.

Moreover, we see that if the elasticity \(\epsilon^H_\phi\) decreases with congestion and \(\epsilon^G_\phi\) increases with congestion, e.g., data traffic is mainly for text content, the sufficient condition in Corollary \ref{cor:welfare maximization congestion responses} must be established and the peering schemes under the welfare-optimal strategies must be the pure free peering. This result is consistent with our intuitions: the paid peering type that maintains a lower congestion level requires more capacity than the settlement-free peering type under the same data load. However, it only brings a small increase of data usage for the congestion-insensitive text traffic. Thus, from the perspective of usage and welfare maximizations, it is worthless to offer the paid peering type for the IAP with limited total capacity, i.e., the pure free peering scheme is better than the hybrid peering scheme. Conversely, if the elasticity \(\epsilon^H_\phi\) increases with congestion and \(\epsilon^G_\phi\) decreases with congestion, e.g., data traffic is mostly for online video, the sufficient condition in Corollary \ref{cor:welfare maximization congestion responses} may not hold and the peering schemes under the welfare-optimal strategies may be the hybrid peering.

\vspace{0.03in}
\textbf{Summary of Implications}: The theoretical results in this subsection could help regulators to make regulatory policies on IAPs' peering schemes. In particular, to protect user welfare, regulators should discourage IAPs to only offer the paid peering type (by Theorem \ref{the:welfare maximization}). Since some low-value CPs have to exit the network platform under the pure paid peering scheme and users would lose the data usage and welfare generated from these CPs.
Furthermore, when network traffic is mostly for text (video), the peering schemes under the welfare-optimal strategies are often to offer only the settlement-free peering type (both the paid and settlement-free peering types) (by Corollary \ref{cor:welfare maximization congestion responses}); 
however, the peering schemes under the profit-optimal strategies are often to offer both the paid and settlement-free peering types (only the paid peering type) (by Corollary \ref{cor:profit maximization congestion responses}). 
The discrepancies suggest that regulators might want to encourage IAPs to allocate more capacity to the settlement-free peering type.

%% file: sections/Simulation.tex
\section{Sensitivities of Optimal Strategies}\label{sec:simulation}

In the previous section, we analyzed the profit-optimal and welfare-optimal strategies, especially their corresponding peering schemes under certain network conditions. With the rapid development of the Internet, characteristics of the market participants, i.e., users, IAPs, and CPs, are continuously changing. For instance, as online video streaming grows fast, users generate higher data demand on CPs and become more sensitive to network congestion, and IAPs are adopting new technologies, e.g., 4G and 5G, to update their capacity.
In this section, we study the sensitivities (dynamics) of the optimal strategies under these varying characteristics of the market participants.

\subsection{Setup of Model Parameters}

We first extend the demand distribution function \(F_w(w)\) and the gain function \(G(\phi)\) to characterize the varying characteristics of the market participants. To capture the growing data demand, we adopt a family of demand distribution function parameterized by \(\alpha\): \(F_w(w,\alpha) = w^\alpha\) for \(0\le w\le 1, \alpha>0\). Under this polynomial function form, as the parameter \(\alpha\) becomes larger, users' data demands are leaning towards higher values. To capture the increasing congestion sensitivity, we choose a family of gain function parameterized by \(\beta\): \(G(\phi,\beta) = 1 - \phi^{\frac{1}{\beta}}\) for \(0\le \phi\le 1, \beta>0\), which satisfies \(G(\phi,\beta_1) > G(\phi,\beta_2)\) for any \(\beta_1<\beta_2\). The parameter \(\beta\) can be regarded as a metric of users' congestion sensitivity, i.e., under the same congestion level, when the parameter \(\beta\) is larger, users' usage gain is smaller and they are more sensitive to network congestion.

\begin{figure}[t]
 \centering
 \includegraphics[width=0.19\textwidth]{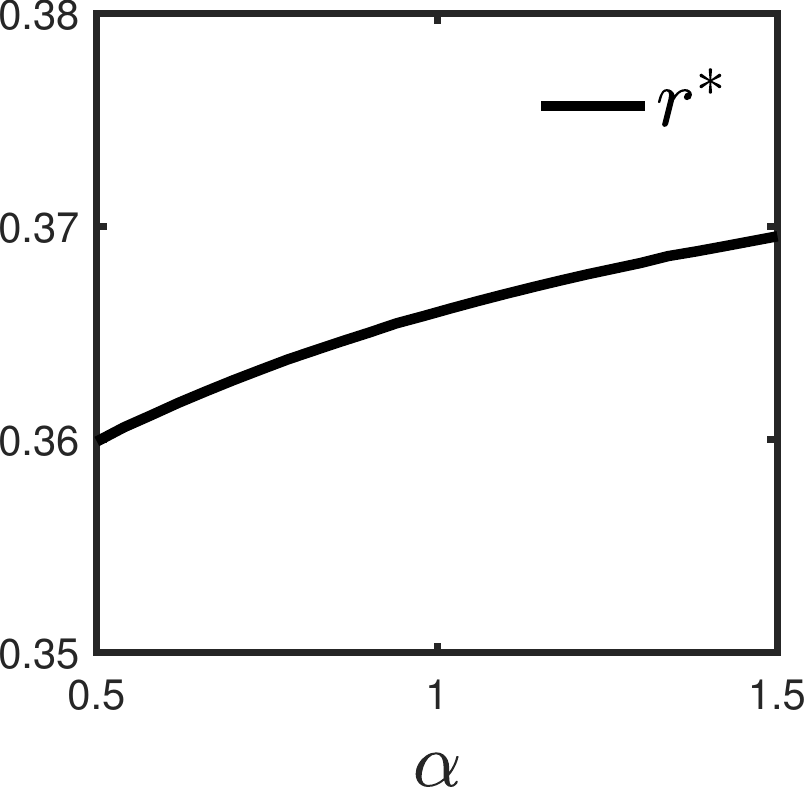}
 \includegraphics[width=0.19\textwidth]{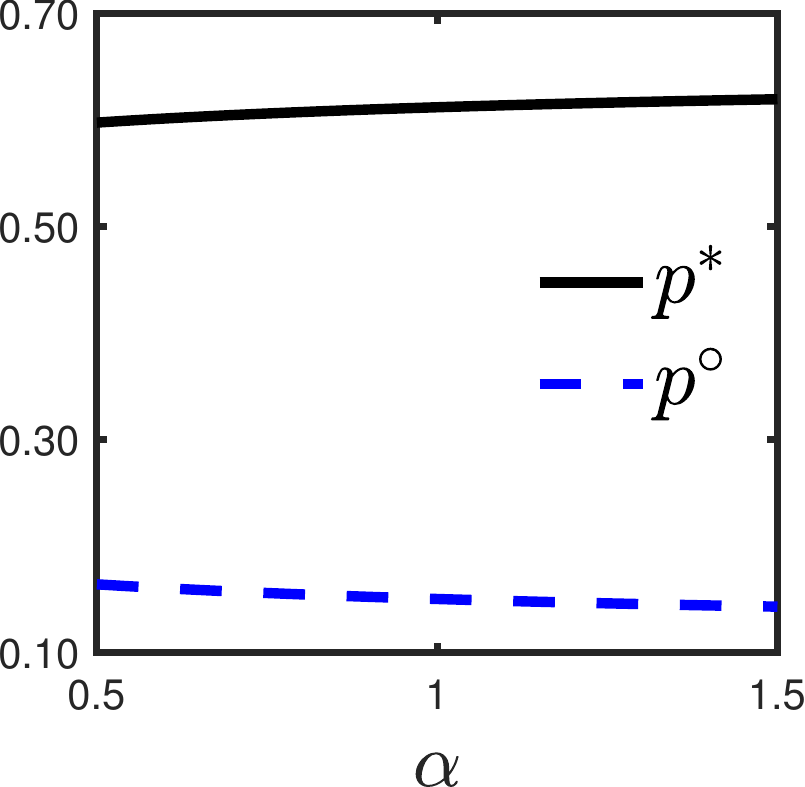}
 \includegraphics[width=0.19\textwidth]{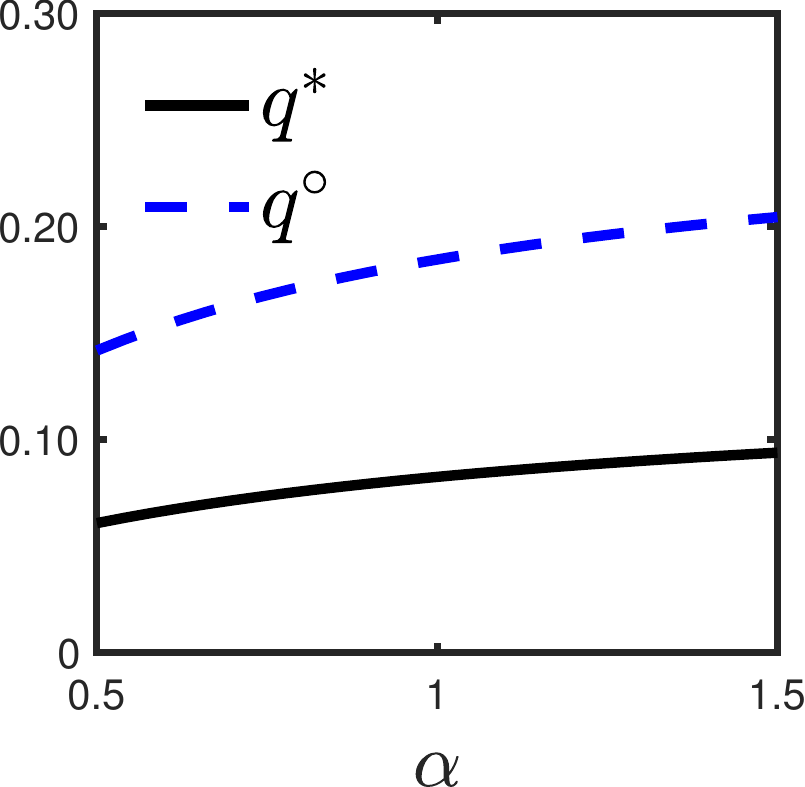}
 \caption{Optimal strategies under varying demand \(\alpha\) when \(\beta = 1, c=0.2\).}
 \label{figure:demand}
\end{figure}
\begin{figure}[t]
 \centering
 \includegraphics[width=0.19\textwidth]{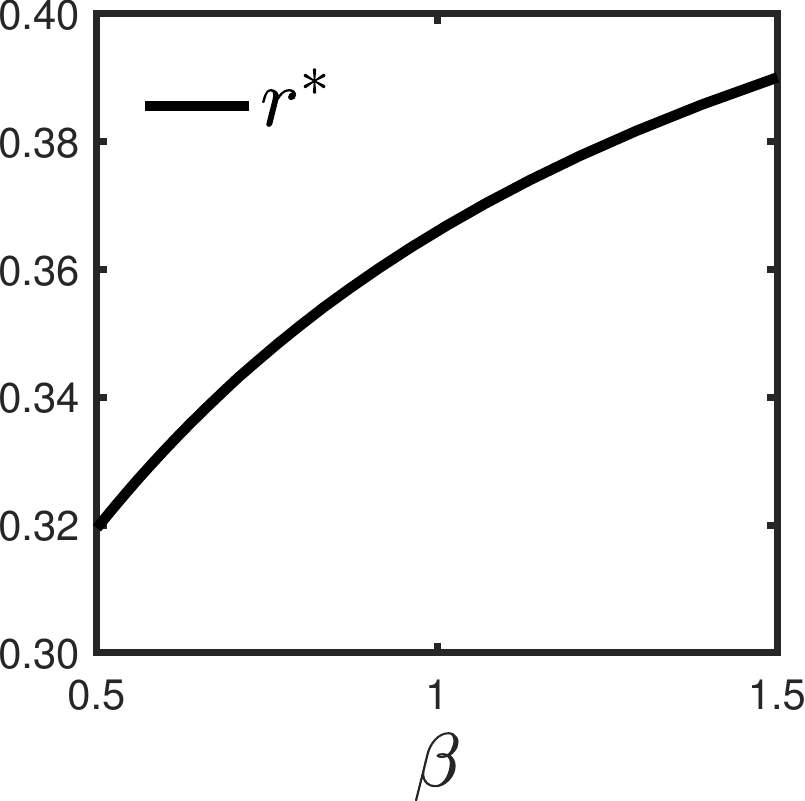}
 \includegraphics[width=0.19\textwidth]{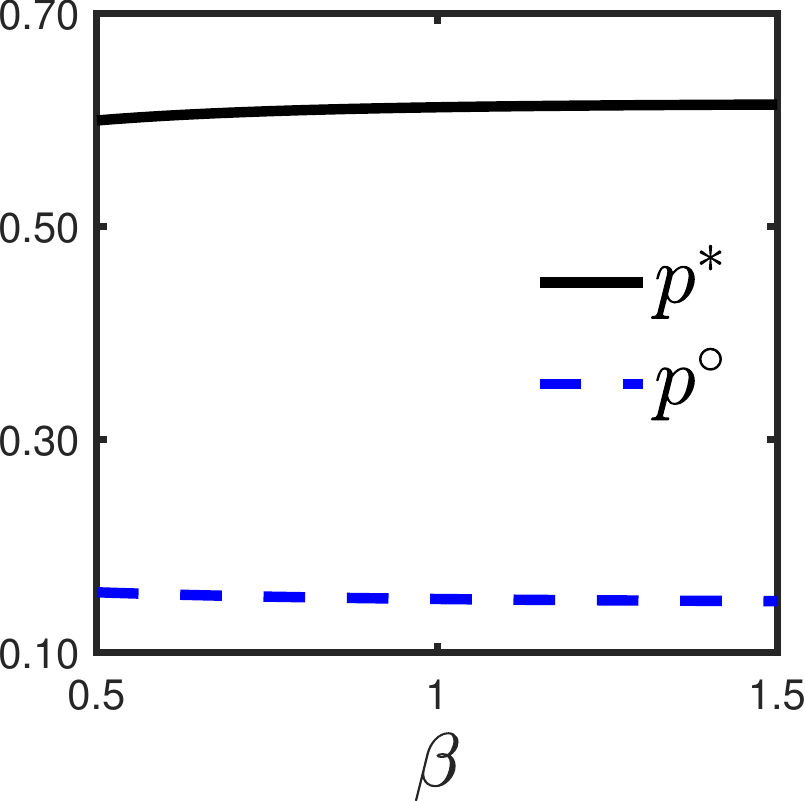}
 \includegraphics[width=0.19\textwidth]{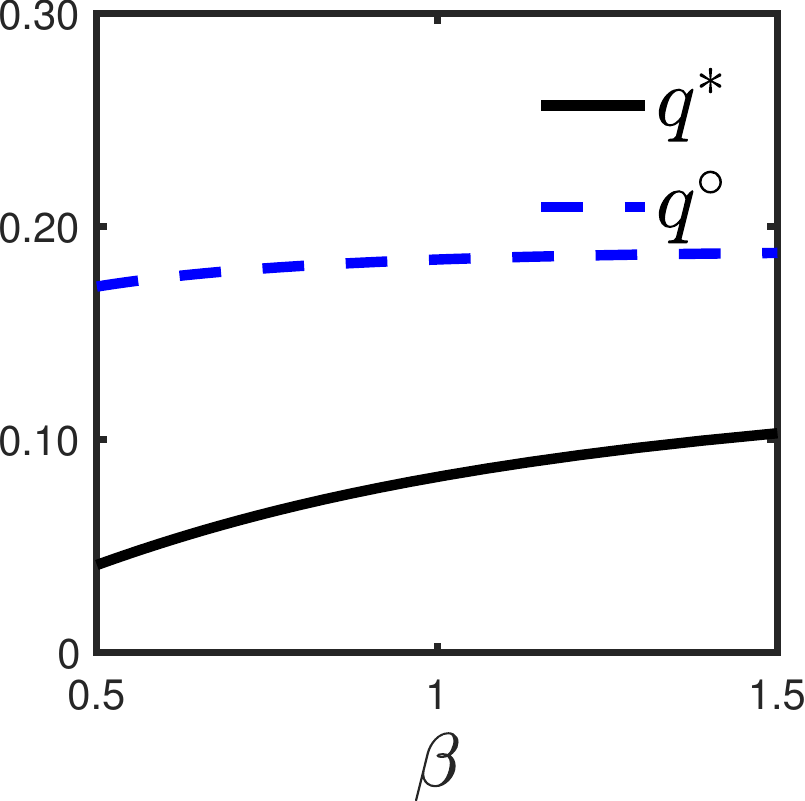}
 \caption{Optimal strategies under varying sensitivity \(\beta\) when \(\alpha = 1, c=0.2\).}
 \label{figure:sensitivity}
\end{figure}

\begin{figure}[t]
 \centering
 \includegraphics[width=0.19\textwidth]{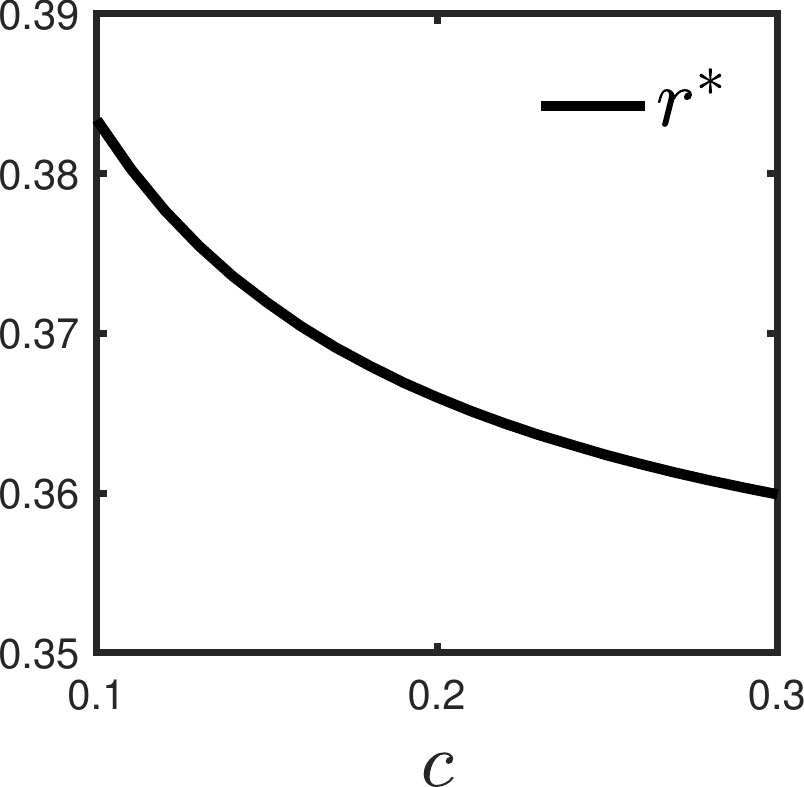}
 \includegraphics[width=0.19\textwidth]{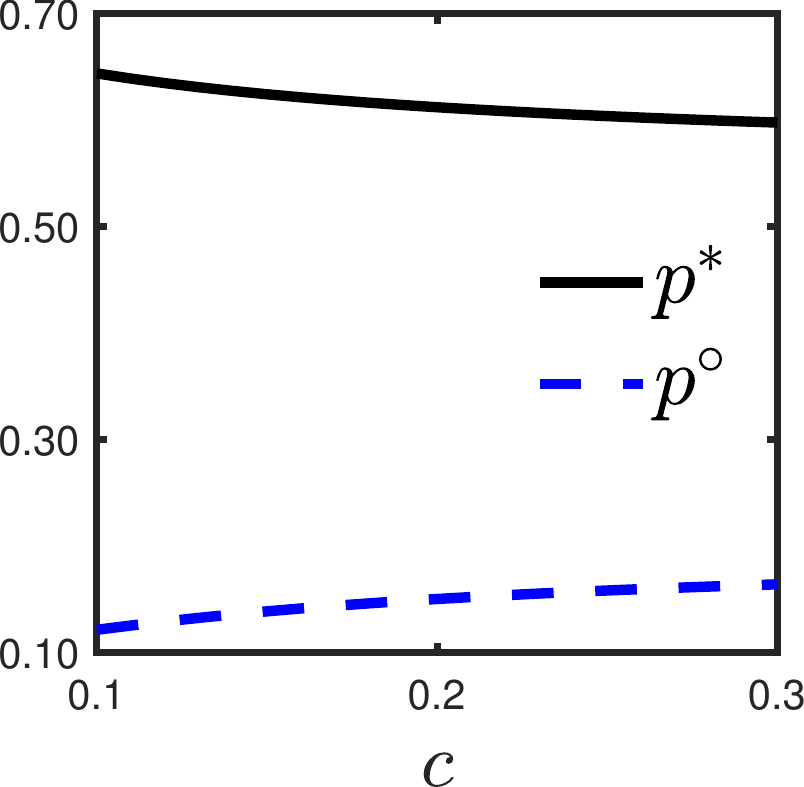}
 \includegraphics[width=0.19\textwidth]{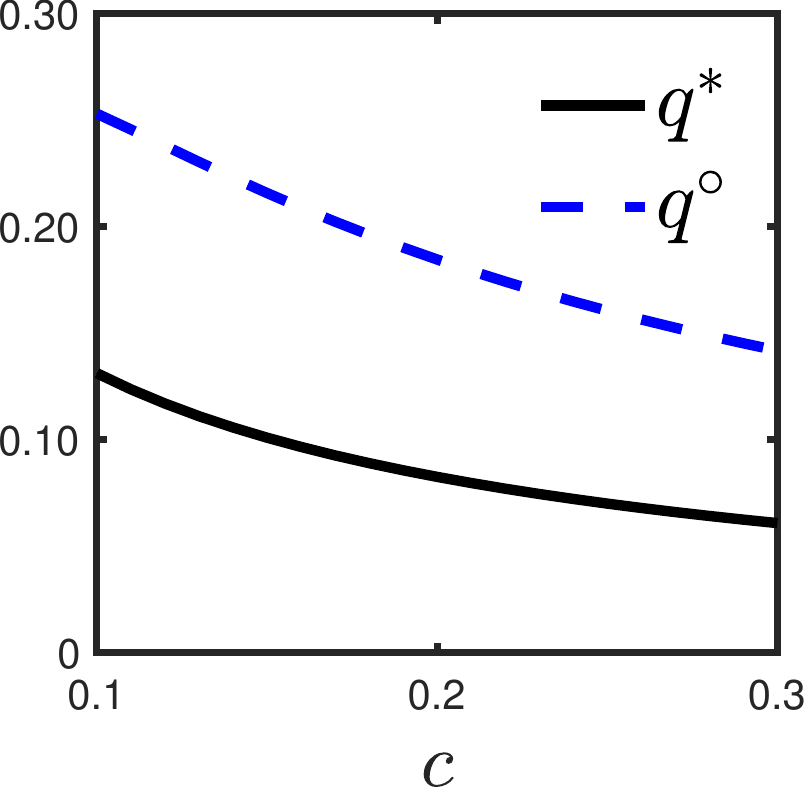}
 \caption{Optimal strategies under varying capacity \(c\) when \(\alpha=\beta=1\).}
 \label{figure:capacity}
\end{figure}

We then choose the forms of the capacity function \(H(\phi)\) and the value distribution functions of users \(F_u(u)\) and CPs \(F_v(v)\). In particular, we use the capacity function \(H(\phi) = 1/\phi\) for \(\phi>0\), under which the required capacity is inversely proportional to the congestion level \(\phi\). This function form was widely used in prior work \cite{gibbens2000internet,jain2001analysis,richard2014pay}. We adopt the value distribution functions \(F_u(u) = u^{0.33}\) and \(F_v(v) = v^{0.33}\) for \(0\le u,v \le 1\), under which the values of users and CPs are leaning towards low values. Besides, we set the cost \(k\) of per-unit data traffic to be \(0.2\).

\subsection{Profit-optimal Strategy}

Under any given parameter of demand distribution \(\alpha\), congestion sensitivity of users \(\beta\), and capacity of the IAP \(c\), we denote the optimal strategy that maximizes the IAP's profit by \(\theta^* = (p^*,q^*,r^*)\).
Next, we explore how this profit-optimal strategy \(\theta^*\) changes with the parameters \(\alpha,\beta\), and \(c\).

Figures \ref{figure:demand}, \ref{figure:sensitivity}, and \ref{figure:capacity} plot the capacity allocation \(r^*\) and prices \(p^*,q^*\) as functions (solid curves) of the parameter of demand distribution \(\alpha\), congestion sensitivity \(\beta\), and IAP's capacity \(c\), respectively.
From Figures \ref{figure:demand} and \ref{figure:sensitivity}, we observe that as \(\alpha\) or \(\beta\) increases, \(r^*\), \(p^*\), and \(q^*\) all increase. 
This observation indicates that when users have larger data demand or higher congestion sensitivity, under the profit-optimal strategy, more capacity would be allocated to the paid peering type and the prices of user side and paid peering type would be higher.
Intuitively, as users become more sensitive to network congestion, the IAP should add the capacity allocated to the paid peering type which maintains a lower congestion level (better data delivery quality) than the settlement-free peering type. The increase of congestion sensitivity also makes the network service more valuable for users and CPs; and therefore, the IAP should raise the prices for extracting more revenue from them.
From Figure \ref{figure:capacity}, we observe that larger values of the capacity \(c\) induce lower values of \(r^*\), \(p^*\), and \(q^*\). This observation implies that when the IAP's capacity is expanded, the fraction of the capacity allocated to the paid peering type and the prices of user side and paid peering type under the profit-optimal strategy would all decrease. Intuitively, if the IAP extend its capacity, the supply of data traffic becomes more abundant and thus the profit-optimal prices would decrease as the basic economic principles of demand and supply implies.

\subsection{Welfare-optimal Strategy}

To protect user welfare, the IAP's strategy usually needs to be regulated. There are two points to note when making regulatory policies. 
First, it is unwise to directly regulate the capacity allocation of the IAP which is often not visible to the public, and the regulatory policies should focus on the prices of user side and paid peering type.
Second, to ensure the feasibility of the policies, the IAP should not incur a loss under the regulated strategy.
Owing to these, we consider welfare-optimal strategies with two constraints: 1) the capacity allocation is profit-optimal, i.e., \(r = r^*\), and 2) the profit of the IAP is always positive, i.e., \(U>0\). Under these constraints, we denote the welfare-optimal prices of user side and paid peering type by \(p^\circ\) and \(q^\circ\), respectively. Next, we study the relationship between the welfare-optimal prices \(p^\circ, q^\circ\) and the profit-optimal prices \(p^*, q^*\).

The second and third subfigures of Figures \ref{figure:demand} to \ref{figure:capacity} plot the prices \(p^\circ\) and \(q^\circ\) as functions (dotted curves) of the parameter of demand distribution \(\alpha\), congestion sensitivity \(\beta\), and IAP's capacity \(c\). 
We observe that the curves of the welfare-optimal prices \(p^\circ\) and \(q^\circ\) are always lower and higher than those of the profit-optimal prices \(p^*\) and \(q^*\), respectively. 
This observation indicates that when the objective changes from profit maximization to welfare maximization, one should lower the price of user side but raise the price of paid peering type. Thus, to protect user welfare, regulators might limit the user-side price and encourage IAPs to change their profit source from the user side to the CP side.
When comparing the trends between the welfare-optimal and profit-optimal prices, we further observe that \(p^\circ\) and \(q^\circ\) have the opposite and same trends as \(p^*\) and \(q^*\), respectively, when \(\alpha\), \(\beta\), or \(c\) varies. In particular, the user-side prices \(p^\circ\) and \(p^*\) decreases and increases with \(\alpha\) or \(\beta\), but increases and decreases with \(c\), respectively. This observation suggests that to protect user welfare, when users' data demand or congestion sensitivity increases, regulators might want to tighten the price regulation on the user side; however, when IAPs expand their capacities, the price regulation might be relaxed moderately.

\vspace{0.03in}
\textbf{Summary of Implications}: The simulation results in this section could help IAPs and regulators to adjust the strategies and the corresponding regulatory policies under varying characteristics of the market participants.
In particular, with the rapid growth of online video streaming, data demand and congestion sensitivity of users are ever-increasing. For optimizing the profits, IAPs should allocate more capacity to the paid peering type and raise the prices of user side and paid peering type. For protecting user welfare, regulators might tighten the price regulation on the user side.
However, when IAPs expand their network capacities, IAPs and regulators should take the opposite operations.

%% file: sections/RelatedWork.tex
\section{Related Work}
Some previous work \cite{reed2014current,lodhi2014open,lodhi2015complexities,ma2017pay,faratin2007complexity} has studied the peering agreements and their impacts on the market participants and evolution of the Internet.
Ma \cite{ma2017pay} adopted a choice model to analyze the peering strategies between IAPs and CPs and showed that whether CPs choose paid peering to connect with IAPs depends on their user stickiness and market shares.
Lodhi {\em at al.} \cite{lodhi2014open,lodhi2015complexities} built an agent-based model to evaluate the peering decisions of transit and tier-2 providers. In \cite{lodhi2015complexities}, they also explored sources of complexity in peering and limitations for tier-2 providers on accurately forecasting the effect of peering decisions.
Reed {\em at al.} \cite{reed2014current} captured the evolution of peering and interconnection to understand the changing characteristics of network traffic. 
Unlike these efforts, our work study the IAP's capacity allocation between the paid and settlement-free peering types, based on which we characterize the IAP's optimal peering schemes that maximize its profit or user welfare.

In our work, as the IAP simultaneously offers the paid and settlement-free peering types, its pricing mechanism on the CP side is a type of Paris Metro Pricing (PMP) \cite{odlyzko1999paris}, which has been studied and used by some prior work \cite{chau2010viability,ma2013public}.
Chau {\em at al.} \cite{chau2010viability} provided sufficient conditions for the viability of the PMP based on a general model of congestion externality. 
Ma and Misra \cite{ma2013public} used the PMP to capture the pricing schemes of non-neutral ISPs and claimed that introducing a public option ISP always benefits end-users.
Both of the work focused on one-sided pricing models, i.e., they only consider the charges of IAPs on end-users \cite{chau2010viability} or CPs \cite{ma2013public}. In the current Internet, however, many IAPs charge both of end-users and CPs. To reflect this state quo, we build a more general two-sided pricing model \cite{Rochet2003}, under which IAPs use the PMP on the CP side and pure usage-based pricing on the user side.


%% file: sections/Conclusions.tex
\section{Conclusions}
In the paper, we model a network platform under which an IAP can offer CPs the paid and settlement-free peering to choose from and charge CPs of using paid peering and end-users.
We capture the data delivery qualities of the peering types by their levels of network congestion and derive an endogenous system congestion in an equilibrium. 
Based on the model, we characterize the IAP's peering schemes under the profit-optimal and welfare-optimal strategies.
We find that the profit and welfare objectives always drive the IAP to offer the paid and settlement-free peering, respectively.
However, whether to simultaneously offer the other peering type largely depends on the type of data traffic, e.g., text or video.
In particular, when data traffic is mostly for text (video), the IAP might (might not) want to offer the settlement-free peering for maximizing its profit and should not (should) offer the paid peering for optimizing user welfare.
This result suggests that regulators might want to encourage the IAP to allocate more capacity to the settlement-free peering.
We also explore the changes of the optimal strategies under varying data demand and congestion sensitivity of users and capacity of the IAP. 
We find that with growing data demand and congestion sensitivity, the IAP needs to allocate more capacity to the paid peering under the profit-optimal strategies. However, when the IAP expands the capacity, it should take the opposite operation.

%% file: sections/Appendix.tex
\appendix
\hspace{-0.13in}\textbf{Proof of Theorem \ref{the:uniqueness}:} We first prove the existence of equilibrium. 
We start with two lemmas.
\begin{lemma}\label{lem1}
For any fixed \(p,q\in (0,+\infty)\) and \(t\in [0,1]\), there exists a unique congestion \(\bm{\phi} = (\phi_h,\phi_l)\) satisfying 
\begin{equation}\label{eq:t_equilibrium}
G(\phi_l) = tG(\phi_h)\quad \text{and}\quad D_h(p,q,\bm{\phi})H(\phi_h)+D_l(p,q,\bm{\phi})H(\phi_l) = c.
\end{equation}
\end{lemma}
Proof of Lemma \ref{lem1}: We define a function \(Y(x) = G^{-1}\big(tG(x)\big)\) for \(x\in [0,1]\). By Assumption \ref{ass:gain}, \(Y(x)\) is non-decreasing in \(x\).
We define a function \(Z(x) = D_h\big(p,q,(x,Y(x))\big)H(x)+D_l\big(p,q,(x,Y(x))\big)H\big(Y(x)\big) - c\) for \(x\in (0,1]\). We denote \(N\triangleq \int_0^{+\infty} w dF_w\). By Equation (\ref{eq:congestion load}), it satisfies that
\begin{align*}
Z(x) &= M(p)N\bigg[1 - F_v\Big(\underline{v}\big(q,(x,Y(x))\big)\Big)\bigg]G(x)H(x) + M(p)NF_v\Big(\underline{v}\big(q,(x,Y(x))\big)\Big)G\big(Y(x)\big)H\big(Y(x)\big) - c\\
&= M(p)N\big[1 - F_v(\frac{q}{1-t})\big]G(x)H(x) + M(p)NF_v(\frac{q}{1-t})G\big(Y(x)\big)H\big(Y(x)\big) - c.
\end{align*}
By Assumption \ref{ass:gain} and \ref{ass:capacity}, we know that \(Z(x)\) is continuously decreasing in \(x\). Furthermore, because \(H(x)\rightarrow +\infty\) as \(x\rightarrow 0\) and \(G(1) = 0\), we have that \(Z(x)\rightarrow +\infty\) as \(x\rightarrow 0\) and \(Z(1) = -c<0\). By the intermediate value theorem, the equation \(Z(x) = 0\) must have a unique solution on the interval \((0,1]\), which we denote by \(x_0\). Then we know that \(\big(x_0,Y(x_0)\big)\) is the unique solution of Equation (\ref{eq:t_equilibrium}). Therefore, Lemma \ref{lem1} is proved.\\

Based on Lemma \ref{lem1}, for any fixed \(p,q,\) and \(t\), we use \(\bm{\phi}(p,q,t) =\big(\phi_h(p,q,t),\phi_l(p,q,t)\big)\) to denote the unique congestion that solves Equation (\ref{eq:t_equilibrium}). Furthermore, we define a function \(R(p,q,t) \triangleq D_h\big(p,q,\bm{\phi}(p,q,t)\big)H\big(\phi_h(p,q,t)\big)/c\).

\begin{lemma}\label{lem2}
For any fixed \(p,q\in (0,+\infty)\), \(R(p,q,t)\) is decreasing in \(t\) and satisfies that \(R(p,q,0)=1\) and \(R(p,q,1)=0\).
\end{lemma}
Proof of Lemma \ref{lem2}: We first show that \(R(p,q,t)\) is decreasing in \(t\) by contradiction. Suppose \(R(p,q,t)\) is not decreasing in \(t\), there must exist \(0<t_1<t_2<1\) satisfying \(R(p,q,t_1)\le R(p,q,t_2)\). By Equation (\ref{eq:t_equilibrium}), we can deduce 
\begin{equation*}
\begin{cases}
\displaystyle\frac{R(p,q,t_1)c}{M(p)NG\big(\phi_h(p,q,t_1)\big)H\big(\phi_h(p,q,t_1)\big)} = 1- F_v(\frac{q}{1-t_1}) > 1- F_v(\frac{q}{1-t_2})= \frac{R(p,q,t_2)c}{M(p)NG\big(\phi_h(p,q,t_2)\big)H\big(\phi_h(p,q,t_2)\big)} \vspace{0.05in}\\
\displaystyle\frac{\big[1-R(p,q,t_1)\big]c}{M(p)NG\big(\phi_l(p,q,t_1)\big)H\big(\phi_l(p,q,t_1)\big)} = F_v(\frac{q}{1-t_1}) < F_v(\frac{q}{1-t_2})= \frac{\big[1-R(p,q,t_2)\big]c}{M(p)NG\big(\phi_l(p,q,t_2)\big)H\big(\phi_l(p,q,t_2)\big)}
\end{cases}
\end{equation*}
implying that \(G\big(\phi_h(p,q,t_1)\big) < G\big(\phi_h(p,q,t_2)\big)\) and \(G\big(\phi_l(p,q,t_1)\big) > G\big(\phi_l(p,q,t_2)\big)\). Thus, by Equation (\ref{eq:t_equilibrium}), we have \(t_1 = G\big(\phi_l(p,q,t_1)\big) /G\big(\phi_h(p,q,t_1)\big) > G\big(\phi_l(p,q,t_2)\big) /G\big(\phi_h(p,q,t_2)\big) = t_2\) 
which is contradictory with the supposition. Therefore, \(R(p,q,t)\) is decreasing in \(t\). 

We then show that \(R(p,q,0)=1\) and \(R(p,q,1)=0\). Because \(G\big(\phi_l(p,q,0)\big) = 0\), we have \(D_l\big(p,q,\bm{\phi}(p,q,0)\big) = 0\). By Equation (\ref{eq:t_equilibrium}), we have \(R(p,q,0) =1\). Because \(D_h\big(p,q,\bm{\phi}(p,q,1)\big) = 0\), we have \(R(p,q,1) =0\). Therefore, Lemma \ref{lem2} is proved.\\

Based on Lemma \ref{lem2}, we denote the reverse function of \(R(p,q,t)\) with respect to \(t\) by \(T(p,q,r) \triangleq R^{-1}(p,q,r)\) for \(r\in [0,1]\). We know that \(T(p,q,r)\) is decreasing in \(r\) and satisfies \(T(p,q,0)=1\) and \(T(p,q,1) = 0\). 

Combining Lemma \ref{lem1} and \ref{lem2}, for any fixed prices \(p,q\), total capacity \(c\) and allocation decision \(r\), we can derive
\begin{equation*}
D_h\Big(p,q,\bm{\phi}\big(p,q,T(p,q,r)\big)\Big)H\Big(\phi_h\big(p,q,T(p,q,r)\big)\Big) \!=\! rc\ \ \text{and} \ \ D_l\Big(p,q,\bm{\phi}\big(p,q,T(p,q,r)\big)\Big)H\Big(\phi_l\big(p,q,T(p,q,r)\big)\Big) \!=\! (1-r)c
\end{equation*}
implying that
\begin{equation*}
\begin{cases}
D_h\Big(p,q,\bm{\phi}\big(p,q,T(p,q,r)\big)\Big)=C^{-1}\Big(rc,\phi_h\big(p,q,T(p,q,r)\big)\Big)\vspace{0.05in}\\
D_l\Big(p,q,\bm{\phi}\big(p,q,T(p,q,r)\big)\Big) \!=\!C^{-1}\Big((1-r)c,\phi_l\big(p,q,T(p,q,r)\big)\Big).
\end{cases}
\end{equation*}
By Definition \ref{def:congestion equilibrium}, \(\bm{\phi}\big(p,q,T(p,q,r)\big)\) is an equilibrium.\\

We next prove the uniqueness of equilibrium by contradiction. By Equation (\ref{eq:congestion load}), we have
\begin{equation}\label{eq:data load expression}
D_h(p,q,\bm{\phi}) = M(p)N \Big[1 - F_v\big(\underline{v}(q,\bm{\phi})\big)\Big] G(\phi_h) \quad \text{and} \quad D_l(p,q,\bm{\phi}) = M(p)N F_v\big(\underline{v}(q,\bm{\phi})\big) G(\phi_l).
\end{equation}
By Assumption \ref{ass:capacity} and Definition \ref{def:congestion equilibrium}, any equilibrium \(\bm{\phi}\) satisfies that
\begin{equation}\label{eq:equilibrium expression}
M(p)N \Big[1 - F_v\big(\underline{v}(q,\bm{\phi})\big)\Big] G(\phi_h)H(\phi_h) = rc \quad\text{and}\quad M(p)N F_v\big(\underline{v}(q,\bm{\phi})\big) G(\phi_l)H(\phi_l) = (1-r)c.
\end{equation}

We prove the uniqueness of equilibrium under three cases: \(r\in (0,1)\), \(r=0\), and \(r=1\). 

For this first case, i.e., \(r\in (0,1)\), suppose there exist two equilibria \(\bm{\hat{\phi}} = (\hat{\phi}_h,\hat{\phi}_l)\) and \(\bm{\bar{\phi}}=(\bar{\phi}_h,\bar{\phi}_l)\). we consider two situations: (i) \(G(\hat{\phi}_l)/G(\hat{\phi}_h) = G(\bar{\phi}_l)/G(\bar{\phi}_h)\) and (ii) \(G(\hat{\phi}_l)/G(\hat{\phi}_h) \neq G(\bar{\phi}_l)/G(\bar{\phi}_h)\). For the situation (i), we have \(\underline{v}(q,\bm{\hat{\phi}}) = \underline{v}(q,\bm{\bar{\phi}})\) by Equation (\ref{eq:gain ratio}). We suppose that \(\hat{\phi}_h<\bar{\phi}_h\) without loss of generality. 
By Equation (\ref{eq:equilibrium expression}), we have
\begin{align*} 
M(p)N \Big[1 - F_v\big(\underline{v}(q,\bm{\hat{\phi}})\big)\Big] G(\hat{\phi}_h)H(\hat{\phi}_h) = rc = M(p)N \Big[1 - F_v\big(\underline{v}(q,\bm{\bar{\phi}})\big)\Big] G(\bar{\phi}_h)H(\bar{\phi}_h)
\end{align*}
implying that \(G(\hat{\phi}_h)H(\hat{\phi}_h) = G(\bar{\phi}_h)H(\bar{\phi}_h)\).
By Assumption \ref{ass:gain} and \ref{ass:capacity}, it satisfies that \(\hat{\phi}_h = \bar{\phi}_h\). This is contradictory with the supposition of \(\hat{\phi}_h<\bar{\phi}_h\). Therefore, the uniqueness of equilibrium is proved for the situation (i). For the situation (ii), we suppose that \(G(\hat{\phi}_l)/G(\hat{\phi}_h) < G(\bar{\phi}_l)/G(\bar{\phi}_h)\) without loss of generality. By Equation (\ref{eq:gain ratio}), we have \(\underline{v}(q,\bm{\hat{\phi}}) < \underline{v}(q,\bm{\bar{\phi}})\). By Equation (\ref{eq:equilibrium expression}), we have
\begin{equation*}
\begin{cases} 
G(\hat{\phi}_h)H(\hat{\phi}_h) =  \displaystyle \frac{rc}{M(p)N \Big[1 - F_v\big(\underline{v}(q,\bm{\hat{\phi}})\big)\Big]} < \frac{rc}{M(p)N \Big[1 - F_v\big(\underline{v}(q,\bm{\bar{\phi}})\big)\Big]} = G(\bar{\phi}_h)H(\bar{\phi}_h)\vspace{0.1in}\\
G(\hat{\phi}_l)H(\hat{\phi}_l) =  \displaystyle \frac{(1-r)c}{M(p)N F_v\big(\underline{v}(q,\bm{\hat{\phi}})\big)} > \frac{(1-r)c}{M(p)N F_v\big(\underline{v}(q,\bm{\bar{\phi}})\big)} = G(\bar{\phi}_l)H(\bar{\phi}_l).
\end{cases}
\end{equation*}
By Assumption \ref{ass:gain} and \ref{ass:capacity}, it satisfies that \(G(\hat{\phi}_h)<G(\bar{\phi}_h)\) and \(G(\hat{\phi}_l)>G(\bar{\phi}_l)\). Thus, we have \(G(\hat{\phi}_l)/G(\hat{\phi}_h) > G(\bar{\phi}_l)/G(\bar{\phi}_h)\) which is contradictory with the supposition of \(G(\hat{\phi}_l)/G(\hat{\phi}_h) < G(\bar{\phi}_l)/G(\bar{\phi}_h)\). Therefore, the uniqueness of equilibrium is proved for the situation (ii).

For the second case, i.e., \(r=0\), any equilibrium \(\bm{\phi}\) must satisfy \(\phi_h = \phi_l\). Otherwise, by Equation (\ref{eq:gain ratio}) and (\ref{eq:data load expression}), it satisfies \(F_v\big(\underline{v}(q,\bm{\phi})\big)<1\) and thus \(D_h(p,q,\bm{\phi}) > C^{-1}(rc,\phi_h) = rc/H(\phi_h) = 0\). By Definition \ref{def:congestion equilibrium}, \(\bm{\phi}\) must not be an equilibrium. 
Suppose there exist two equilibria \(\bm{\hat{\phi}} = (\hat{\phi}_h,\hat{\phi}_l)\) and \(\bm{\bar{\phi}}=(\bar{\phi}_h,\bar{\phi}_l)\) satisfying \(\hat{\phi}_h = \hat{\phi}_l\) and \(\bar{\phi}_h = \bar{\phi}_l\). By Equation (\ref{eq:gain ratio}), we know that \(F_v\big(\underline{v}(q,\bm{\hat{\phi}})\big) = F_v\big(\underline{v}(q,\bm{\bar{\phi}})\big) = 1\). By Equation (\ref{eq:equilibrium expression}), we have that
\[M(p)N G(\hat{\phi}_l)H(\hat{\phi}_l) = (1-r)c = M(p)N G(\bar{\phi}_l)H(\bar{\phi}_l).\]
By Assumption \ref{ass:gain} and \ref{ass:capacity}, we know that \(\hat{\phi}_l = \bar{\phi}_l\). Thus, \(\bm{\hat{\phi}}\) and \(\bm{\bar{\phi}}\) are the same point which is contradictory with the supposition.

For the third case, i.e., \(r=1\), any equilibrium \(\bm{\phi}\) must satisfy \(\phi_l = 1\). Otherwise, by Assumption \ref{ass:gain} and Equation (\ref{eq:data load expression}), it satisfies \(G(\phi_l)>0\) and thus \(D_l(p,q,\bm{\phi}) > C^{-1}\big((1-r)c,\phi_l) = (1-r)c/H(\phi_l) = 0\). By Definition \ref{def:congestion equilibrium}, \(\bm{\phi}\) must not be an equilibrium. 
Suppose there exist two equilibria \(\bm{\hat{\phi}} = (\hat{\phi}_h,\hat{\phi}_l)\) and \(\bm{\bar{\phi}}=(\bar{\phi}_h,\bar{\phi}_l)\) satisfying \(\hat{\phi}_l =  \bar{\phi}_l = 1\). By Equation (\ref{eq:gain ratio}), we know that \(F_v\big(\underline{v}(q,\bm{\hat{\phi}})\big) = F_v\big(\underline{v}(q,\bm{\bar{\phi}})\big) = F_v(q)\). By Equation (\ref{eq:equilibrium expression}), we have that
\[M(p)N \big[1-F_v(q)\big]G(\hat{\phi}_h)H(\hat{\phi}_h) = rc = M(p)N\big[1-F_v(q)\big] G(\bar{\phi}_h)H(\bar{\phi}_h).\]
By Assumption \ref{ass:gain} and \ref{ass:capacity}, we know that \(\hat{\phi}_h = \bar{\phi}_h\). Thus, \(\bm{\hat{\phi}}\) and \(\bm{\bar{\phi}}\) are the same point which is contradictory with the supposition. Therefore, the uniqueness of equilibrium is proved.
\QEDA\\

\hspace{-0.13in}\textbf{Proof of Corollary \ref{cor:pc effect}:} We first prove that \(\varphi_h\) is decreasing in \(p\) by contradiction. Suppose \(\varphi_h\) is not decreasing in \(p\), there must exist \(\theta_1=(p_1,q,r)\) and \(\theta_2 = (p_2,q,r)\) satisfying \(p_1<p_2\) and \(\varphi_h (\theta_1,c) \le \varphi_h (\theta_2,c)\). By Assumption \ref{ass:gain} and \ref{ass:capacity}, we have \(G\big(\varphi_h (\theta_1,c)\big) \ge G\big(\varphi_h (\theta_2,c)\big)\) and \(H\big(\varphi_h (\theta_1,c)\big) \ge H\big(\varphi_h (\theta_2,c)\big)\). By Equation (\ref{eq:equilibrium expression}), we have
\begin{equation*}
1 - F_v\Big(\underline{v}\big(q,\bm{\varphi}(\theta_1,c)\big)\Big) = \frac{rcN^{-1}}{M(p_1)G\big(\varphi_h (\theta_1,c)\big)H\big(\varphi_h (\theta_1,c)\big)} < \frac{rcN^{-1}}{M(p_2)G\big(\varphi_h (\theta_2,c)\big)H\big(\varphi_h (\theta_2,c)\big)} = 1 - F_v\Big(\underline{v}\big(q,\bm{\varphi}(\theta_2,c)\big)\Big)
\end{equation*}
implying that \(\underline{v}\big(q,\bm{\varphi}(\theta_1,c)\big) > \underline{v}\big(q,\bm{\varphi}(\theta_2,c)\big)\). By Equation (\ref{eq:gain ratio}), we can derive 
\[\frac{G\big(\varphi_l (\theta_1,c)\big)}{G\big(\varphi_l (\theta_2,c)\big)} > \frac{G\big(\varphi_h (\theta_1,c)\big)}{G\big(\varphi_h (\theta_2,c)\big)}\ge 1\]
implying that \(\varphi_l (\theta_1,c) < \varphi_l (\theta_2,c)\).
Furthermore, By Equation (\ref{eq:equilibrium expression}), we have
\begin{equation*}
F_v\Big(\underline{v}\big(q,\bm{\varphi}(\theta_1,c)\big)\Big)= \frac{(1-r)cN^{-1}}{M(p_1)G\big(\varphi_l (\theta_1,c)\big)H\big(\varphi_l (\theta_1,c)\big)} < \frac{(1-r)cN^{-1}}{M(p_2)G\big(\varphi_l (\theta_2,c)\big)H\big(\varphi_l (\theta_2,c)\big)} = F_v\Big(\underline{v}\big(q,\bm{\varphi}(\theta_2,c)\big)\Big)
\end{equation*}
implying that \(\underline{v}\big(q,\bm{\varphi}(\theta_1,c)\big) < \underline{v}\big(q,\bm{\varphi}(\theta_2,c)\big)\). This is contradictory with the result of \(\underline{v}\big(q,\bm{\varphi}(\theta_1,c)\big) > \underline{v}\big(q,\bm{\varphi}(\theta_2,c)\big)\). Therefore, \(\varphi_h\) must be decreasing in \(p\). Furthermore, because \(d_h = C^{-1}(rc,\varphi_h) = rc/H(\varphi_h)\) by Definition \ref{def:congestion equilibrium} and \(H(\varphi_h)\) is decreasing in \(\varphi_h\) by Assumption \ref{ass:capacity}, we know that \(d_h\) is decreasing in \(p\). Similarly, we can prove that \(\varphi_l\) and \(d_l\) are decreasing in \(p\).
\QEDA\\

\hspace{-0.13in}\textbf{Proof of Corollary \ref{cor:q effect}:} We first prove that \(\varphi_l\) is increasing in \(q\) by contradiction. Suppose \(\varphi_l\) is not increasing in \(q\), there must exist \(\theta_1=(p,q_1,r)\) and \(\theta_2 = (p,q_2,r)\) satisfying \(q_1<q_2\) and \(\varphi_l (\theta_1,c) \ge \varphi_l (\theta_2,c)\). By Assumption \ref{ass:gain} and \ref{ass:capacity}, we have \(G\big(\varphi_l (\theta_1,c)\big) \le G\big(\varphi_l (\theta_2,c)\big)\) and \(H\big(\varphi_l (\theta_1,c)\big) \le H\big(\varphi_l (\theta_2,c)\big)\). By Equation (\ref{eq:equilibrium expression}), it satisfies that
\begin{equation*}
F_v\Big(\underline{v}\big(q_1,\bm{\varphi}(\theta_1,c)\big)\Big)= \frac{(1-r)cN^{-1}}{M(p)G\big(\varphi_l (\theta_1,c)\big)H\big(\varphi_l (\theta_1,c)\big)} \ge \frac{(1-r)cN^{-1}}{M(p)G\big(\varphi_l (\theta_2,c)\big)H\big(\varphi_l (\theta_2,c)\big)} = F_v\Big(\underline{v}\big(q_2,\bm{\varphi}(\theta_2,c)\big)\Big)
\end{equation*}
implying that \(\underline{v}\big(q_1,\bm{\varphi}(\theta_1,c)\big) \ge \underline{v}\big(q_2,\bm{\varphi}(\theta_2,c)\big)\). By Equation (\ref{eq:gain ratio}), we can derive 
\[\frac{G\big(\varphi_h (\theta_1,c)\big)}{G\big(\varphi_h (\theta_2,c)\big)}<\frac{G\big(\varphi_l (\theta_1,c)\big)}{G\big(\varphi_l (\theta_2,c)\big)}\le 1\]
implying that \(\varphi_h (\theta_1,c) > \varphi_h (\theta_2,c)\).
Furthermore, By Equation (\ref{eq:equilibrium expression}), we have
\begin{equation*}
1 - F_v\Big(\underline{v}\big(q_1,\bm{\varphi}(\theta_1,c)\big)\Big)= \frac{rcN^{-1}}{M(p)G\big(\varphi_h (\theta_1,c)\big)H\big(\varphi_h (\theta_1,c)\big)} > \frac{rcN^{-1}}{M(p)G\big(\varphi_h (\theta_2,c)\big)H\big(\varphi_h (\theta_2,c)\big)} = 1 - F_v\Big(\underline{v}\big(q_2,\bm{\varphi}(\theta_2,c)\big)\Big)
\end{equation*}
implying that \(\underline{v}\big(q_1,\bm{\varphi}(\theta_1,c)\big) < \underline{v}\big(q_2,\bm{\varphi}(\theta_2,c)\big)\). This is contradictory with the result of \(\underline{v}\big(q_1,\bm{\varphi}(\theta_1,c)\big) \ge \underline{v}\big(q_2,\bm{\varphi}(\theta_2,c)\big)\). Therefore, \(\varphi_l\) must be increasing in \(p\). Furthermore, because \(d_l = C^{-1}\big((1-r)c,\varphi_l\big) = (1-r)c/H(\varphi_l)\) by Definition \ref{def:congestion equilibrium} and \(H(\varphi_l)\) is decreasing in \(\varphi_l\) by Assumption \ref{ass:capacity}, we know that \(d_l\) is also increasing in \(q\). Similarly, we can prove that \(\varphi_h\) and \(d_h\) are both decreasing in \(q\).
\QEDA\\

\hspace{-0.13in}\textbf{Proof of Corollary \ref{cor:r effect}:} We first prove that \(\varphi_l\) is increasing in \(r\) by contradiction. Suppose \(\varphi_l\) is not increasing in \(r\), there must exist \(\theta_1=(p,q,r_1)\) and \(\theta_2 = (p,q,r_2)\) satisfying \(0<r_1<r_2<1\) and \(\varphi_l (\theta_1,c) \ge \varphi_l (\theta_2,c)\). By Assumption \ref{ass:gain} and \ref{ass:capacity}, we have \(G\big(\varphi_l (\theta_1,c)\big) \le G\big(\varphi_l (\theta_2,c)\big)\) and \(H\big(\varphi_l (\theta_1,c)\big) \le H\big(\varphi_l (\theta_2,c)\big)\). By Equation (\ref{eq:equilibrium expression}), we have
\begin{equation*}
F_v\Big(\underline{v}\big(q,\bm{\varphi}(\theta_1,c)\big)\Big)= \frac{(1-r_1)cN^{-1}}{M(p)G\big(\varphi_l (\theta_1,c)\big)H\big(\varphi_l (\theta_1,c)\big)} > \frac{(1-r_2)cN^{-1}}{M(p)G\big(\varphi_l (\theta_2,c)\big)H\big(\varphi_l (\theta_2,c)\big)} = F_v\Big(\underline{v}\big(q,\bm{\varphi}(\theta_2,c)\big)\Big)\end{equation*}
implying that \(\underline{v}\big(q,\bm{\varphi}(\theta_1,c)\big) > \underline{v}\big(q,\bm{\varphi}(\theta_2,c)\big)\). By Equation (\ref{eq:gain ratio}), we can derive 
\[\frac{G\big(\varphi_h (\theta_1,c)\big)}{G\big(\varphi_h (\theta_2,c)\big)}<\frac{G\big(\varphi_l (\theta_1,c)\big)}{G\big(\varphi_l (\theta_2,c)\big)}\le 1\]
implying that \(\varphi_h (\theta_1,c) > \varphi_h (\theta_2,c)\). Thus, we know that \(G\big(\varphi_h (\theta_1,c)\big) < G\big(\varphi_h (\theta_2,c)\big)\) and \(H\big(\varphi_h (\theta_1,c)\big) < H\big(\varphi_h (\theta_2,c)\big)\) by Assumption \ref{ass:gain} and \ref{ass:capacity}.
Because \(\varphi_h\le \varphi_l\), it satisfies that \(G\big(\varphi_h (\theta_1,c)\big)H\big(\varphi_h (\theta_1,c)\big)\ge G\big(\varphi_l (\theta_1,c)\big)H\big(\varphi_l (\theta_1,c)\big)\). Combining Equation (\ref{eq:equilibrium expression}), we can deduce that
\begin{equation*}
\begin{aligned}
&\frac{\big[r_1+(1-r_1)\big]c}{M(p)N}  = \left[1 - F_v\Big(\underline{v}\big(q,\bm{\varphi}(\theta_1,c)\big)\Big)\right]G\big(\varphi_h (\theta_1,c)\big)H\big(\varphi_h (\theta_1,c)\big) + F_v\Big(\underline{v}\big(q,\bm{\varphi}(\theta_1,c)\big)\Big)G\big(\varphi_l (\theta_1,c)\big)H\big(\varphi_l (\theta_1,c)\big)\\
\le &\Big[1 - F_v\Big(\underline{v}\big(q,\bm{\varphi}(\theta_2,c)\big)\Big)\Big]G\big(\varphi_h (\theta_1,c)\big)H\big(\varphi_h (\theta_1,c)\big) + F_v\Big(\underline{v}\big(q,\bm{\varphi}(\theta_2,c)\big)\Big)G\big(\varphi_l (\theta_1,c)\big)H\big(\varphi_l (\theta_1,c)\big)\\ 
< &\Big[1 - F_v\Big(\underline{v}\big(q,\bm{\varphi}(\theta_2,c)\big)\Big)\Big]G\big(\varphi_h (\theta_2,c)\big)H\big(\varphi_h (\theta_2,c)\big) + F_v\Big(\underline{v}\big(q,\bm{\varphi}(\theta_2,c)\big)\Big)G\big(\varphi_l (\theta_2,c)\big)H\big(\varphi_l (\theta_2,c)\big) = \frac{\big[r_2+(1-r_2)\big]c}{M(p)N} 
\end{aligned}
\end{equation*}
which must not hold. Therefore, \(\varphi_l\) must be increasing in \(r\). Similarly, we can prove that \(d_l\) and \(d_h\) are increasing and decreasing in \(r\), respectively.

We then prove that \(\varphi_l= \varphi_h\) if and only if \(r=0\). If \(r=0\), by Equation (\ref{eq:equilibrium expression}), we have
\[M(p)N \Big[1 - F_v\big(\underline{v}(q,\bm{\varphi})\big)\Big] G(\varphi_h) H(\varphi_h) = rc = 0\]
implying that \(F_v\big(\underline{v}(q,\bm{\varphi})\big)= 1\). By Equation (\ref{eq:gain ratio}), we know that \(\varphi_l = \varphi_h\). Conversely, we can deduce \(r=0\) if \(\varphi_l = \varphi_h\).

Finally, we prove that \(\varphi_l = 1\) if and only if \(r=1\). If \(r=1\), by Equation (\ref{eq:equilibrium expression}), we have
\[M(p)N F_v\big(\underline{v}(q,\bm{\varphi})\big) G(\varphi_l) H(\varphi_l) = (1-r)c = 0\]
implying that \(G(\varphi_l) = 0\) and thus \(\varphi_l = 1\) by Assumption \ref{ass:gain}. Conversely, we can deduce \(r=1\) if \(\varphi_l = 1\).
\QEDA\\

\hspace{-0.13in}\textbf{Proof of Theorem \ref{the:profit maximization}:} We first prove that the capacity allocation decision of any profit-optimal strategy must not be zero by contradiction. Suppose there exists a profit-optimal strategy \(\theta^* = (p^*,q^*,0)\). Because the capacity \(c\) and cost \(k\) are constants, we denote \(\varphi_h(\theta) \triangleq \varphi_h(\theta,c), \varphi_l(\theta) \triangleq \varphi_l(\theta,c), d_h(\theta) \triangleq d_h(\theta,c), d_l(\theta) \triangleq d_l(\theta,c),\) and \(U(\theta) \triangleq U(\theta,k,c)\) for simplicity. 
Because \(\theta^*\) is a global maximum of the profit function \(U(\theta)\), we have \(p^*+q^*>k\). Otherwise, \(U(\theta^*)  = (p^*+ q^* -k) d_h(\theta^*) + (p^*-k)d_l(\theta^*) \le 0\) and thus \(\theta^*\) must not be a global maximum of the profit function \(U\). By Karush-Kuhn-Tucker (KKT) necessary conditions, we also have that
\begin{equation}\label{eq:the2_1}
\frac{\partial U(\theta)}{\partial r}\bigg|_{\theta = \theta^*} \le 0.
\end{equation}
By Corollary \ref{cor:r effect}, it satisfies that
\[\frac{\partial d_h(\theta)}{\partial r}\bigg|_{\theta = \theta^*} > 0 \quad \text{and} \quad \frac{\partial d_l(\theta)}{\partial r}\bigg|_{\theta = \theta^*} < 0.\]
If \(p^*\le k\), we can deduce that
\begin{equation}\label{eq:the2_2}
\frac{\partial U(\theta)}{\partial r}\bigg|_{\theta = \theta^*}  = (p^*+ q^* -k)\frac{\partial d_h(\theta)}{\partial r}\bigg|_{\theta = \theta^*} + (p^*-k)\frac{\partial d_l(\theta)}{\partial r}\bigg|_{\theta = \theta^*}> 0
\end{equation}
which is contradictory with Equation (\ref{eq:the2_1}). Therefore, we know that \(p^* > k\).

We define a triple \(\eta \triangleq (p,q,t)\) for \(p,q\in (0,+\infty)\) and \(t\in [0,1]\). Combining Lemma \ref{lem1} and \ref{lem2}, we know that \(\bm{\phi}(\eta) = \bm{\varphi}\big(p,q,R(\eta)\big)\). We denote \(d_h(\eta) \triangleq d_h\big(p,q,R(\eta)\big), d_l(\eta) \triangleq d_l\big(p,q,R(\eta)\big),\) and \(U(\eta) \triangleq U\big(p,q,R(\eta)\big)\). Because it satisfies \(T(p,q,0)=1\) for any prices \(p,q\), we can denote \(\eta^* \triangleq \big(p^*,q^*,T(\theta^*)\big) = (p^*,q^*,1)\). If \(\theta^*\) is a global maximum of the profit function \(U(\theta)\), \(\eta^*\) is a global maximum of the profit function \(U(\eta)\), i.e., it solves the optimization problem:
\begin{align*}
\underset{\eta}{\text{maximize}} \quad & U(\eta) = (p+q-k) d_h(\eta) + (p-k) d_l(\eta) \\
\text{subject to} \quad & \eta\in (0,+\infty) \times (0,+\infty) \times [0,1].
\end{align*}
We define a function \(\underline{V}\big(q,t\big) \triangleq q/(1-t)\) which is increasing in \(t\). By Equation (\ref{eq:gain ratio}), it satisfies that
\[\underline{v}\big(q,\bm{\phi(\eta)}\big) =  \frac{qG\big(\phi_h(\eta)\big)}{G\big(\phi_h(\eta)\big)-G\big(\phi_l(\eta)\big)} =  \frac{q}{1-t} =  \underline{V}\big(q,t\big).\]
Combining Equation (\ref{eq:equilibrium expression}), we have the identity 
\[d_h(\eta) H\big(\varphi_h(\eta)\big) \!+\! d_l(\eta)H\big(\varphi_l(\eta)\big) \!=\!  M(p)N\Big\{\big[1-F_v(\underline{V}(q,t))\big]G\big(\varphi_h(\eta)\big)H\big(\varphi_h(\eta)\big) \!+\! F_v\big(\underline{V}(q,t)\big)G\big(\varphi_l(\eta)\big)H\big(\varphi_l(\eta)\big)\Big\} \!=\! c\]
for any \(\eta\), from which we can deduce that
\begin{equation}\label{heta}
\begin{cases}
\displaystyle\frac{\partial \phi_h(\eta)}{\partial p} = \frac{\phi_h}{K(\eta)\epsilon^G_\phi(\phi_h)}\frac{M'(p)}{M(p)}\Big[d_h(\eta)H(\phi_h)+d_l(\eta)H(\phi_l)\Big]\vspace{0.05in}\\
\displaystyle\frac{\partial \phi_h(\eta)}{\partial q} = \frac{\phi_h}{K(\eta)\epsilon^G_\phi(\phi_h)}\frac{\partial F_v\big(\underline{V}(q,t)\big)}{\partial q}\Big[\frac{d_l(\eta)H(\phi_l)}{F_v\big(\underline{V}(q,t)\big)}-\frac{d_h(\eta)H(\phi_h)}{1-F_v\big(\underline{V}(q,t)\big)}\Big]\vspace{0.05in}\\
\displaystyle\frac{\partial \phi_h(\eta)}{\partial t} = \frac{\phi_h}{K(\eta)\epsilon^G_\phi(\phi_h)}\left\{\frac{\partial F_v\big(\underline{V}(q,t)\big)}{\partial t}\Big[\frac{d_l(\eta)H(\phi_l)}{F_v\big(\underline{V}(q,t)\big)}-\frac{d_h(\eta)H(\phi_h)}{1-F_v\big(\underline{V}(q,t)\big)}\Big]+d_l(\eta)\Big[\frac{H(\phi_l)}{t}+ H'(\phi_l)\frac{G(\phi_h)}{G'(\phi_l)}\Big]\right\}
\end{cases}
\end{equation}
where \(\phi_h = \phi_h(\eta)\), \(\phi_l = \phi_l(\eta)\), and the label \(K(\eta)\) is defined by
\[K(\eta) \triangleq d_hH(\phi_h)\left(\frac{\epsilon^H_\phi(\phi_h)}{\epsilon^G_\phi(\phi_h)}+1\right)+d_lH(\phi_l)\left(\frac{\epsilon^H_\phi(\phi_l)}{\epsilon^G_\phi(\phi_l)}+1\right).\]
Because it always satisfies that \(G\big(\phi_l(\eta)\big) = t G\big(\phi_h(\eta)\big)\), we have
\begin{equation}\label{leta}
\displaystyle\frac{\partial \phi_l(\eta)}{\partial p} = \frac{\phi_l\epsilon^G_\phi(\phi_h)}{\phi_h\epsilon^G_\phi(\phi_l)}\frac{\partial \phi_h(\eta)}{\partial p}, \quad \frac{\partial \phi_l(\eta)}{\partial q} = \frac{\phi_l\epsilon^G_\phi(\phi_h)}{\phi_h\epsilon^G_\phi(\phi_l)}\frac{\partial \phi_h(\eta)}{\partial q}, \quad \text{and} \quad \frac{\partial \phi_l(\eta)}{\partial t} = \frac{\phi_l\epsilon^G_\phi(\phi_h)}{\phi_h\epsilon^G_\phi(\phi_l)}\frac{\partial \phi_h(\eta)}{\partial t} + \frac{G(\phi_h)}{G'(\phi_l)}.
\end{equation}
By Equation (\ref{heta}) and (\ref{leta}), we can further derive that
\begin{equation}\label{eq:the2_4}
\begin{cases}
\displaystyle\frac{\partial d_h(\eta)}{\partial p}  = \frac{d_hM'(p)}{K(\eta)M(p)}\left[\frac{d_hH(\phi_h)\epsilon^H_\phi(\phi_h)}{\epsilon^G_\phi(\phi_h)}+\frac{d_lH(\phi_l)\epsilon^H_\phi(\phi_l)}{\epsilon^G_\phi(\phi_l)}\right]\le 0\vspace{0.05in}\\
\displaystyle\frac{\partial d_l(\eta)}{\partial p}  = \frac{d_lM'(p)}{K(\eta)M(p)}\left[\frac{d_hH(\phi_h)\epsilon^H_\phi(\phi_h)}{\epsilon^G_\phi(\phi_h)}+\frac{d_lH(\phi_l)\epsilon^H_\phi(\phi_l)}{\epsilon^G_\phi(\phi_l)}\right]\le 0\vspace{0.05in}\\
\displaystyle\frac{\partial d_h(\eta)}{\partial q}  = \frac{d_h}{K(\eta)}\frac{\partial F_v\big(\underline{V}(q,t)\big)}{\partial q}\left[\frac{d_h H(\phi_h)}{1 - F_v\big(\underline{V}(q,t)\big)} - \frac{d_l H(\phi_l)}{F_v\big(\underline{V}(q,t)\big)} \right] - \frac{\partial F_v\big(\underline{V}(q,t)\big)}{\partial q}\frac{d_h}{1- F_v\big(\underline{V}(q,t)\big)}\le 0\vspace{0.05in}\\
\displaystyle\frac{\partial d_l(\eta)}{\partial q}  = \frac{d_l}{K(\eta)}\frac{\partial F_v\big(\underline{V}(q,t)\big)}{\partial q}\left[\frac{d_h H(\phi_h)}{1 - F_v\big(\underline{V}(q,t)\big)} - \frac{d_l H(\phi_l)}{F_v\big(\underline{V}(q,t)\big)} \right] + \frac{\partial F_v\big(\underline{V}(q,t)\big)}{\partial q}\frac{d_l}{F_v\big(\underline{V}(q,t)\big)}\ge 0 \vspace{0.05in}\\
\displaystyle\frac{\partial d_h(\eta)}{\partial t}  = \frac{\partial d_h(\eta)}{\partial q} \left(\frac{\partial F_v\big(\underline{V}(q,t)\big)}{\partial q}\right)^{-1}\frac{\partial F_v\big(\underline{V}(q,t)\big)}{\partial t}  - \frac{d_h}{tK(\eta)}\left[d_lH(\phi_l)\left(\frac{\epsilon^H_\phi(\phi_l)}{\epsilon^G_\phi(\phi_l)}+1\right)\right]< 0\vspace{0.05in}\\
\displaystyle\frac{\partial d_l(\eta)}{\partial t}  = \frac{\partial d_l(\eta)}{\partial q} \left(\frac{\partial F_v\big(\underline{V}(q,t)\big)}{\partial q}\right)^{-1}\frac{\partial F_v\big(\underline{V}(q,t)\big)}{\partial t}  + \frac{d_l}{tK(\eta)}\left[d_hH(\phi_h)\left(\frac{\epsilon^H_\phi(\phi_h)}{\epsilon^G_\phi(\phi_h)}+1\right)\right]> 0.
\end{cases}
\end{equation}

Because \(\phi_h(\eta),\phi_l(\eta)\) are continuous in \(t\in [0,1]\) and \(\phi_h(\eta) = \phi_l(\eta)\) when \(t =1\), we have \(G\big(\phi_h(\eta)\big)/G\big(\phi_l(\eta)\big) \rightarrow 1\) and \(G\big(\phi_h(\eta)\big)H\big(\phi_h(\eta)\big) - G\big(\phi_l(\eta)\big)H\big(\phi_l(\eta)\big) \rightarrow 0\) as \(t\rightarrow 1\).
By Equation (\ref{eq:the2_4}), it satisfies that
\begin{equation*}
\displaystyle\frac{\partial d_h(\eta)}{\partial q}\left(\displaystyle\frac{\partial d_l(\eta)}{\partial q}\right)^{-1} = \frac{d_h\big[K(\eta)\big]^{-1}\big[G\big(\phi_h(\eta)\big)H\big(\phi_h(\eta)\big) - G\big(\phi_l(\eta)\big)H\big(\phi_l(\eta)\big)\big]- G\big(\phi_h(\eta)\big)}{d_l\big[K(\eta)\big]^{-1}\big[G\big(\phi_h(\eta)\big)H\big(\phi_h(\eta)\big) - G\big(\phi_l(\eta)\big)H\big(\phi_l(\eta)\big)\big]+ G\big(\phi_l(\eta)\big)}\rightarrow -1 \quad \text{as} \quad t \rightarrow 1 
\end{equation*}
from which we can further derive that
\begin{equation*}
\displaystyle\frac{\partial d_h(\eta)}{\partial t}\left(\displaystyle\frac{\partial d_l(\eta)}{\partial t}\right)^{-1} \rightarrow -1 \quad \text{as} \quad t \rightarrow 1. 
\end{equation*}
Thus, given prices \(p^*\) and \(q^*\), there must exist \(t_0\in [0,1)\) satisfying
\begin{equation*}
\displaystyle\frac{\partial d_h(\eta)}{\partial t}\bigg|_{\eta =(p^*,q^*,t_1)}\left(\displaystyle\frac{\partial d_l(\eta)}{\partial t}\bigg|_{\eta =(p^*,q^*,t_1)}\right)^{-1} < -\frac{p^*-k}{p^*+q^*-k}, \quad  \forall \ t_1\in (t_0,1). 
\end{equation*}
Thus we can derive that
\begin{equation}\label{eq:KKTcon}
\displaystyle\frac{\partial U(\eta)}{\partial t}\bigg|_{\eta =(p^*,q^*,t_1)} = (p^*+q^*-k)\frac{\partial d_h(\eta)}{\partial t}\bigg|_{\eta =(p^*,q^*,t_1)} + (p^*- k)\frac{\partial d_l(\eta)}{\partial t}\bigg|_{\eta =(p^*,q^*,t_1)} < 0, \quad  \forall \ t_1\in (t_0,1).
\end{equation}
Because \(\eta^*\) is a global maximum of the profit function \(U(\eta)\) and \(U(\eta)\) is continuous in \(t\in [t_0,1]\), by the Lagrange Mean Value Theorem, there exists \(t_2\in (t_0,1)\) satisfying
\[\frac{\partial U(\eta)}{\partial t}\bigg|_{\eta =(p^*,q^*,t_2)} = \frac{U(\eta^*)-U(p^*,q^*,t_0)}{1 - t_0}\ge 0\]
which is contradictory with Equation (\ref{eq:KKTcon}). Therefore, \(\eta^*\) must not be a global maximum of the profit function \(U(\eta)\) and thus \(\theta^*=(p^*,q^*,0)\) must not be a global maximum of the profit function \(U(\theta)\). In other word, the capacity allocation decision of any profit-optimal strategy must not be zero.

We then prove that Equation (\ref{eq:profit maximization average}) and (\ref{eq:profit maximization ratio}) hold. If a strategy \(\theta = (p,q,r)\) is the profit-optimal, it must be a global maximum of the profit function \(U(\theta)\). By KKT necessary conditions,
we have 
\begin{equation*}
\begin{cases}
\displaystyle\frac{\partial U(\theta)}{\partial p} = (p+q-k) \frac{\partial d_h(\theta)}{\partial p} + d_h(\theta)+ (p-k)\frac{\partial d_l(\theta)}{\partial p} + d_l(\theta) = -\frac{(p+q-k)d_h}{p}\epsilon^{d_h(\theta)}_{p} - \frac{(p-k)d_l}{p}\epsilon^{d_l(\theta)}_{p}  + d_t = 0\\
\displaystyle\frac{\partial U(\theta)}{\partial q} = (p+q-k) \frac{\partial d_h(\theta)}{\partial q} + d_h(\theta)+ (p-k)\frac{\partial d_l(\theta)}{\partial q}  = -\frac{(p+q-k)d_h}{q}\epsilon^{d_h(\theta)}_{q} - \frac{(p-k)d_l}{q}\epsilon^{d_l(\theta)}_{q}  + d_h = 0
\end{cases}
\end{equation*}
from which we can derive that
\begin{equation*}
\begin{cases}
(p+q-k)d_h\epsilon^{d_h}_p + (p - k) d_l \epsilon^{d_l}_p =  p d_t\vspace{0.05in}\\
\displaystyle\frac{(p+q-k)d_h}{(p-k)d_l} = \frac{-pd_t\epsilon^{d_l}_q+qd_h\epsilon^{d_l}_p}{pd_t\epsilon^{d_h}_q-qd_h\epsilon^{d_h}_p}.
\end{cases}
\end{equation*}
By KKT necessary conditions, we also have that
\begin{equation*}
\displaystyle\frac{\partial U(\theta)}{\partial r} = (p+q-k) \frac{\partial d_h(\theta)}{\partial r} + (p-k)\frac{\partial d_l(\theta)}{\partial r}   =  -\frac{(p+q-k)d_h}{r}\epsilon^{d_h(\theta)}_{r} - \frac{(p-k)d_l}{r}\epsilon^{d_l(\theta)}_{r} 
\begin{cases}
=0 & \text{if} \ \ r\in (0,1);\vspace{0.05in}\\
\ge 0 & \text{if} \ \ r = 1
\end{cases}
\end{equation*}
from which we can deduce that
\begin{equation*}
\frac{(p+q-k)d_h}{(p-k)d_l} 
\begin{cases}
=-\displaystyle\frac{\epsilon^{d_l}_r}{\epsilon^{d_h}_r} & \text{if} \ \ r\in (0,1);\vspace{0.05in}\\
\ge -\displaystyle\frac{\epsilon^{d_l}_r}{\epsilon^{d_h}_r} &  \text{if} \  \ r = 1.
\end{cases}
\end{equation*}
Therefore, Equation (\ref{eq:profit maximization average}) and (\ref{eq:profit maximization ratio}) hold.
\QEDA\\

\hspace{-0.13in}\textbf{Proof of Corollary \ref{cor:profit maximization congestion responses}:} We prove that the capacity allocation decision of any profit-optimal strategy must be one by contradiction. Suppose there exists a profit-optimal strategy \(\theta^* = (p^*,q^*,r^*)\) satisfying \(r^*\neq 1\). Combining Theorem \ref{the:profit maximization}, we know \(r^*\in (0,1)\).
Because \(\theta^*\) is a global maximum of the profit function \(U(\theta)\), we have \(p^*+q^*>k\). Otherwise, \(U(\theta^*)  = (p^*+ q^* -k) d_h(\theta^*) + (p^*-k)d_l(\theta^*) \le 0\) and thus \(\theta^*\) must not be a global maximum of the profit function \(U\). By Karush-Kuhn-Tucker (KKT) necessary conditions, we also have that
\begin{equation}\label{eq:supply1}
\frac{\partial U(\theta)}{\partial r}\bigg|_{\theta = \theta^*} = 0.
\end{equation}
By Corollary \ref{cor:r effect}, it satisfies that
\[\frac{\partial d_h(\theta)}{\partial r}\bigg|_{\theta = \theta^*} > 0 \quad \text{and} \quad \frac{\partial d_l(\theta)}{\partial r}\bigg|_{\theta = \theta^*} < 0.\]
If \(p^*\le k\), we can deduce that
\begin{equation}\label{eq:supply2}
\frac{\partial U(\theta)}{\partial r}\bigg|_{\theta = \theta^*}  = (p^*+ q^* -k)\frac{\partial d_h(\theta)}{\partial r}\bigg|_{\theta = \theta^*} + (p^*-k)\frac{\partial d_l(\theta)}{\partial r}\bigg|_{\theta = \theta^*}> 0
\end{equation}
which is contradictory with Equation (\ref{eq:supply1}). Therefore, we know that \(p^* > k\).
We define \(\eta^* \triangleq \big(p^*,q^*,T(\theta^*)\big)\). Because \(\theta^*\) is a global maximum of the profit function \(U(\theta)\), \(\eta^*\) is a global maximum of the profit function \(U(\eta)\). Because \(r^*\in (0,1)\), we have \(T(\theta^*)\in (0,1)\). By KKT necessary conditions, we have that
\begin{align}
&\displaystyle\frac{\partial U(\eta)}{\partial q}\bigg|_{\eta = \eta^*} = (p^* + q^* - k)\frac{\partial d_h(\eta)}{\partial q}\bigg|_{\eta = \eta^*} + (p^* - k)\frac{\partial d_l(\eta)}{\partial q}\bigg|_{\eta = \eta^*} + d_h(\eta^*) = 0.\label{eq:cor4_1}\vspace{0.05in}\\
&\displaystyle\frac{\partial U(\eta)}{\partial t}\bigg|_{\eta = \eta^*} = (p^* + q^* - k)\frac{\partial d_h(\eta)}{\partial t}\bigg|_{\eta = \eta^*} + (p^* - k)\frac{\partial d_l(\eta)}{\partial t}\bigg|_{\eta = \eta^*} = 0.\label{eq:cor4_2}
\end{align}
For any \(p,q\in (0,+\infty)\) and \(t\in [0,1]\), we define a function of the triple \(\eta \triangleq (p,q,t)\) by
\begin{equation}\label{eq:cor4_3}
L(\eta) = (p+q-k)\left[H\big(\phi_l(\eta)\big)\left(\frac{\epsilon^H_\phi\big(\phi_l(\eta)\big)}{\epsilon^G_\phi\big(\phi_l(\eta)\big)}+1\right)\right] - (p-k)\left[H\big(\phi_h(\eta)\big)\left(\frac{\epsilon^H_\phi\big(\phi_h(\eta)\big)}{\epsilon^G_\phi\big(\phi_h(\eta)\big)}+1\right)\right].
\end{equation}
If \(H(\epsilon^H_\phi/\epsilon^G_\phi+1)\) is an increasing function of the congestion level \(\phi\), it satisfies that \(L(\eta^*)>0\).
Combining Equation (\ref{eq:the2_4}) and (\ref{eq:cor4_1}), we can deduce that
\begin{align*}
\frac{\partial U(\eta)}{\partial t}\bigg|_{\eta =\eta^*} =& \left[\frac{\partial U(\eta)}{\partial q}\bigg|_{\eta = \eta^*}-d_h(\eta^*)\right]\!\!\left[\frac{\partial F_v\big(\underline{V}(q,t)\big)}{\partial q}\bigg|_{(q,t) = (q^*,T(\theta^*))}\right]^{-1}\!\!\!\frac{\partial F_v\big(\underline{V}(q,t)\big)}{\partial t}\bigg|_{(q,t) = (q^*,T(\theta^*))} \!-\! \frac{ d_h(\eta^*)d_l(\eta^*)L(\eta^*)}{T(\theta^*)K(\eta^*)}\vspace{0.05in}\\
< & \left[\frac{\partial U(\eta)}{\partial q}\bigg|_{\eta = \eta^*}-d_h(\eta^*)\right]\!\!\left[\frac{\partial F_v\big(\underline{V}(q,t)\big)}{\partial q}\bigg|_{(q,t) = (q^*,T(\theta^*))}\right]^{-1}\!\!\!\frac{\partial F_v\big(\underline{V}(q,t)\big)}{\partial t}\bigg|_{(q,t) = (q^*,T(\theta^*))}\vspace{0.05in}\\
=& \left[\frac{\partial U(\eta)}{\partial q}\bigg|_{\eta = \eta^*}-d_h(\eta^*)\right] \frac{q^*}{1-T(\theta^*)}<0
\end{align*}
which is contradictory with Equation (\ref{eq:cor4_2}). Therefore, \(\eta^*\) must not be a global maximum of the profit function \(U(\eta)\) and thus \(\theta^*=(p^*,q^*,r^*)\) must not be a global maximum of the profit function \(U(\theta)\). In other word, the capacity allocation decision of any profit-optimal strategy must be one, if \(H(\epsilon^H_\phi/\epsilon^G_\phi+1)\) is an increasing function of the congestion level \(\phi\).
\QEDA\\

\hspace{-0.13in}\textbf{Proof of Corollary \ref{cor:profit maximization hazard rate responses}:} We prove that the capacity allocation decision of any profit-optimal strategy must be in \((0,1)\) by contradiction. Suppose there exists a profit-optimal strategy \(\theta^* = (p^*,q^*,r^*)\) satisfying \(r^*\notin (0,1)\). Combining Theorem \ref{the:profit maximization}, we know \(r^*=1\). We denote \(\eta^* \triangleq \big(p^*,q^*,T(\theta^*)\big) = (p^*,q^*,0)\) and it must be a global maximum of the profit function \(U(\eta)\). By Corollary \ref{cor:r effect}, we have \(\varphi_l(\theta) = 1\) and \(d_l(\theta) = 0\) for any strategy \(\theta =(p,q,1)\). Therefore, we have \(d_l(\eta) = 0\) for any \(\eta = (p,q,0)\) and thus 
\[d_l(\eta^*) = 0 \quad \text{and} \quad \frac{\partial d_l(\eta)}{\partial p}\bigg|_{\eta = \eta^*} = \frac{\partial d_l(\eta)}{\partial q}\bigg|_{\eta = \eta^*} = 0.\]
Furthermore, by KKT necessary conditions, it satisfies that
\begin{equation}\label{eq:cor5_1}
\begin{cases}
\displaystyle\frac{\partial U(\eta)}{\partial p}\bigg|_{\eta = \eta^*} = (p^* + q^* - k)\frac{\partial d_h(\eta)}{\partial p}\bigg|_{\eta = \eta^*}  + d_h(\eta^*) = 0.\vspace{0.05in}\\
\displaystyle\frac{\partial U(\eta)}{\partial q}\bigg|_{\eta = \eta^*} = (p^* + q^* - k)\frac{\partial d_h(\eta)}{\partial q}\bigg|_{\eta = \eta^*}  + d_h(\eta^*)  = 0.
\end{cases}
\end{equation}
Combining Equation (\ref{eq:the2_4}) and (\ref{eq:cor5_1}), we can deduce 
\[\displaystyle\tilde{F}_v(q^*) \!=\! \frac{1}{1- F_v(q^*)}\frac{\partial F_v(q)}{\partial q}\bigg|_{q = q^*} \!\!=\! \frac{1}{1- F_v\big(\underline{V}(q^*,0)\big)}\frac{\partial F_v\big(\underline{V}(q,t)\big)}{\partial q}\bigg|_{(q,t) = (q^*,0)} \!\!=\!\frac{-1}{M(p^*)}\frac{\partial M(p)}{\partial p}\bigg|_{p = p^*} \!\!=\! \frac{F'_u(p^*)}{1-F_u(p^*)} \!=\!  \tilde{F}_u(p^*)\]
which is contradictory with the condition of Corollary \ref{cor:profit maximization hazard rate responses}. Therefore, \(\eta^*\) must not be a global maximum of the profit function \(U(\eta)\) and thus \(\theta^*=(p^*,q^*,r^*)\) must not be a global maximum of the profit function \(U(\theta)\). In other word, the capacity allocation decision of any profit-optimal strategy must be in \((0,1)\), if \(\tilde{F}_u(p) < \tilde{F}_v(q)\) holds for any prices \(p,q\in (0,+\infty).\)
\QEDA\\

\hspace{-0.13in}\textbf{Proof of Theorem \ref{the:welfare maximization}:} We first prove that the capacity allocation decision of any welfare-optimal strategy must not be one by contradiction. Suppose there exists a welfare-optimal strategy \(\theta^\circ = (p^\circ,q^\circ,r^\circ)\) satisfying \(r^\circ = 1\). Because the capacity \(c\) is a constant, we denote \(W(\theta) \triangleq W(\theta,c)\). Because \(r^\circ= 1\), by Corollary \ref{cor:r effect}, we have \(\varphi_l(\theta^\circ) = 1\) and thus
\[d_l(\theta^\circ) = 0 \quad \text{and} \quad \frac{\partial d_l(\theta)}{\partial q}\bigg|_{\theta = \theta^\circ} = 0.\]
Because \(\theta^\circ\) is a global maximum of the welfare function \(W(\theta)\), by KKT necessary conditions, we have
\begin{equation*}
\frac{\partial W(\theta)}{\partial q}\bigg|_{\theta = \theta^\circ} = s(p^\circ) \frac{\partial d_t(\theta)}{\partial q}\bigg|_{\theta = \theta^\circ} = s(p^\circ) \frac{\partial d_h(\theta)}{\partial q}\bigg|_{\theta = \theta^\circ} = 0
\end{equation*}
implying that
\begin{equation}\label{eq:fina}
\frac{\partial d_h(\theta)}{\partial q}\bigg|_{\theta = \theta^\circ} = 0.
\end{equation}
However, we can prove that \(\varphi_h(\theta)\) and \(d_h(\theta)\) are decreasing in \(q\) as \(r=1\) by contradiction as follows. Suppose \(\varphi_h(\theta)\) is not decreasing in \(q\), there must exist \(\theta_1=(p,q_1,1)\) and \(\theta_2 = (p,q_2,1)\) satisfying \(q_1<q_2\) and \(\varphi_h (\theta_1) \le \varphi_h (\theta_2)\). By Assumption \ref{ass:gain} and \ref{ass:capacity}, we have \(G\big(\varphi_h (\theta_1)\big) \ge G\big(\varphi_h (\theta_2)\big)\) and \(H\big(\varphi_h (\theta_1)\big) \ge H\big(\varphi_h (\theta_2)\big)\). By Equation (\ref{eq:equilibrium expression}), it satisfies that
\begin{equation*}
1 - F_v\Big(\underline{v}\big(q_1,\bm{\varphi}(\theta_1)\big)\Big)= \frac{c}{M(p)NG\big(\varphi_h(\theta_1)\big)H\big(\varphi_h (\theta_1)\big)} \le\frac{c}{M(p)NG\big(\varphi_h (\theta_2)\big)H\big(\varphi_h (\theta_2)\big)} = 1 - F_v\Big(\underline{v}\big(q_2,\bm{\varphi}(\theta_2)\big)\Big)
\end{equation*}
implying that \(\underline{v}\big(q_1,\bm{\varphi}(\theta_1)\big) \ge \underline{v}\big(q_2,\bm{\varphi}(\theta_2)\big)\). Because \(\varphi_l(\theta_1) = \varphi_l(\theta_2) = 1\), by Equation (\ref{eq:gain ratio}), we have \(q_1\ge q_2\) which is contradictory with the supposition of \(q_1< q_2\). Thus \(\varphi_h(\theta)\) must be decreasing in \(q\) as \(r=1\). Furthermore, because \(d_h(\theta) = C^{-1}\big(rc,\varphi_h(\theta)\big) = rc/H\big(\varphi_h(\theta)\big)\) by Definition \ref{def:congestion equilibrium} and \(H(\phi)\) is decreasing in \(\phi\) by Assumption \ref{ass:capacity}, we know that \(d_h(\theta)\) is decreasing in \(q\) as \(r=1\). 
This result is contradictory with Equation (\ref{eq:fina}). Therefore, \(\theta^\circ = (p^\circ,q^\circ,r^\circ) = (p^\circ,q^\circ,1)\) must not be a global maximum of the welfare function \(W(\theta)\). In other word, the capacity allocation decision of any welfare-optimal strategy must not be one.

We then prove that Equation (\ref{eq:welfare maximization}) holds.
If a strategy \(\theta = (p,q,r)\) is the welfare-optimal, it must be a global maximum of the welfare function \(W(\theta)\). By KKT necessary conditions,
we have 
\begin{equation*}
\displaystyle\frac{\partial W(\theta)}{\partial p} = \frac{\partial s(p)}{\partial p} d_t(\theta) + s(p) \frac{\partial d_t(\theta)}{\partial p} = 0 \quad \text{and}\quad\frac{\partial W(\theta)}{\partial q} = s(p) \frac{\partial d_t(\theta)}{\partial q} = 0
\end{equation*}
from which we can deduce that
\(\epsilon^s_p + \epsilon^{d_t(\theta)}_p =  0\) and \(\epsilon^{d_t(\theta)}_q = 0\).
By KKT necessary conditions, we also have
\begin{equation*}
\displaystyle\frac{\partial W(\theta)}{\partial r} = s(p) \frac{\partial d_t(\theta)}{\partial r} 
\begin{cases}
=0 & \text{if} \ \ r\in (0,1);\vspace{0.05in}\\
\le 0 & \text{if} \ \ r = 0
\end{cases}
\end{equation*}
from which we can deduce that \(\epsilon^{d_t(\theta)}_r=0\) if \(r\in (0,1)\), and \(\epsilon^{d_t(\theta)}_r\ge 0\) if \(r = 0\).
Therefore, Equation (\ref{eq:welfare maximization}) holds.
\QEDA\\

\hspace{-0.13in}\textbf{Proof of Corollary \ref{cor:welfare maximization congestion responses}:} We prove that the capacity allocation decision of any welfare-optimal strategy must be zero by contradiction. Suppose there exists a welfare-optimal strategy \(\theta^\circ = (p^\circ,q^\circ,r^\circ)\) satisfying that \(r^\circ\neq 0\). Combining Theorem \ref{the:welfare maximization}, we know \(r^\circ\in (0,1)\) and thus \(T(\theta^\circ) \in (0,1)\). We denote \(\eta^\circ \triangleq \big(p^\circ,q^\circ,T(\theta^\circ)\big)\). We denote \(W(\eta) \triangleq W\big(p,q,R(\eta)\big)\) for any triple \(\eta = (p,q,t)\in (0,+\infty)\times (0,+\infty)\times [0,1]\). Because \(\theta^\circ\) is a global maximum of the welfare function \(W(\theta)\), \(\eta^\circ\) is a global maximum of the welfare function \(W(\eta)\). By KKT necessary conditions, it satisfies
\begin{align}
&\displaystyle\frac{\partial W(\eta)}{\partial q}\bigg|_{\eta = \eta^\circ} = s(p^\circ)\left[\frac{\partial d_h(\eta)}{\partial q}\bigg|_{\eta = \eta^\circ} + \frac{\partial d_l(\eta)}{\partial q}\bigg|_{\eta = \eta^\circ} \right]= 0,\label{eq:cor6_1}\vspace{0.05in}\\
&\displaystyle\frac{\partial W(\eta)}{\partial t}\bigg|_{\eta = \eta^\circ} = s(p^\circ)\left[\frac{\partial d_h(\eta)}{\partial t}\bigg|_{\eta = \eta^\circ} + \frac{\partial d_l(\eta)}{\partial t}\bigg|_{\eta = \eta^\circ} \right]= 0.\label{eq:cor6_2}
\end{align}
We define a function
\[J(\eta) = H\big(\phi_l(\eta)\big)\left(\frac{\epsilon^H_\phi\big(\phi_l(\eta)\big)}{\epsilon^G_\phi\big(\phi_l(\eta)\big)}+1\right) - H\big(\phi_h(\eta)\big)\left(\frac{\epsilon^H_\phi\big(\phi_h(\eta)\big)}{\epsilon^G_\phi\big(\phi_h(\eta)\big)}+1\right).\]
If \(H(\epsilon^H_\phi/\epsilon^G_\phi+1)\) is a decreasing function of the congestion level \(\phi\), we have \(J(\eta^\circ)<0\).
By Equation (\ref{eq:the2_4}) and (\ref{eq:cor6_1}), we can derive
\begin{align*}
\frac{\partial W(\eta)}{\partial t}\bigg|_{\eta =\eta^\circ} =& \frac{\partial W(\eta)}{\partial q}\bigg|_{\eta = \eta^\circ}\left[\frac{\partial F_v\big(\underline{V}(q,t)\big)}{\partial q}\bigg|_{(q,t) = (q^\circ,T(\theta^\circ))}\right]^{-1}\!\!\!\frac{\partial F_v\big(\underline{V}(q,t)\big)}{\partial t}\bigg|_{(q,t) = (q^\circ,T(\theta^\circ))}-\frac{s(p^\circ)d_h(\eta^\circ)d_l(\eta^\circ)J(\eta^\circ)}{T(\theta^\circ)K(\eta^\circ)}\vspace{0.05in}\\
> & \frac{\partial W(\eta)}{\partial q}\bigg|_{\eta = \eta^\circ}\left[\frac{\partial F_v\big(\underline{V}(q,t)\big)}{\partial q}\bigg|_{(q,t) = (q^\circ,T(\theta^\circ))}\right]^{-1}\!\!\!\frac{\partial F_v\big(\underline{V}(q,t)\big)}{\partial t}\bigg|_{(q,t) = (q^\circ,T(\theta^\circ))}=0
\end{align*}
which is contradictory with Equation (\ref{eq:cor6_2}). Therefore, \(\eta^\circ\) must not be a global maximum of the welfare function \(W(\eta)\) and thus \(\theta^\circ = (p^\circ,q^\circ,r^\circ)\) must not be a global maximum of the welfare function \(W(\theta)\). In other word, the capacity allocation decision of any welfare-optimal strategy must be zero, if \(H(\epsilon^H_\phi/\epsilon^G_\phi+1)\) is a decreasing function of the congestion level \(\phi\).
\QEDA\\

%% file: report.bbl
\begin{thebibliography}{10}

\bibitem{akamai}
Akamai.
\newblock \url{https://www.akamai.com}.

\bibitem{att}
{AT\&T}.
\newblock \url{https://www.att.com/}.

\bibitem{ComcastXFINITY}
Comcast {X}finity.
\newblock \url{http://www.xfinity.com/}.

\bibitem{facebook}
Facebook.
\newblock \url{https://www.facebook.com/}.

\bibitem{netflix}
Netflix.
\newblock \url{https://www.netflix.com/}.

\bibitem{sprint}
Sprint.
\newblock \url{https://www.sprint.com}.

\bibitem{tmobile}
{T-Mobile}.
\newblock \url{http://www.t-mobile.com/}.

\bibitem{FCC_Open}
The {U}.{S}. {FCC}'s {O}pen {I}nternet {O}rder.
\newblock \url{https://www.fcc.gov/document/fcc-releases-open-internet-order}.

\bibitem{FCC}
{U.S. Federal Communications Commission}.
\newblock \url{https://www.fcc.gov}.

\bibitem{VerizonFios}
Verizon {F}ios.
\newblock \url{http://www.verizon.com/}.

\bibitem{comcastusage}
Your next comcast bill may be priced per gigabyte.
\newblock
  \url{http://fortune.com/2015/09/30/comcast-broadband-pricing-wireless/}.

\bibitem{chau2010viability}
C.-K. Chau, Q.~Wang, and D.-M. Chiu.
\newblock {On the viability of Paris Metro Pricing for communication and
  service networks}.
\newblock In {\em Proceedings of IEEE INFOCOM}, pages 1--9, 2010.

\bibitem{faratin2007complexity}
P.~Faratin, D.~D. Clark, S.~Bauer, and W.~Lehr.
\newblock Complexity of {I}nternet interconnections: Technology, incentives and
  implications for policy.
\newblock In {\em TPRC conference}, 2007.

\bibitem{gibbens2000internet}
R.~Gibbens, R.~Mason, and R.~Steinberg.
\newblock {{I}nternet service classes under competition}.
\newblock {\em IEEE Journal on Selected Areas in Communications},
  18(12):2490--2498, 2000.

\bibitem{hande10pricing}
P.~Hande, M.~Chiang, R.~Calderbank, and J.~Zhang.
\newblock Pricing under constraints in access networks: Revenue maximization
  and congestion management.
\newblock In {\em Proceedings of IEEE INFOCOM}, pages 1--9, 2010.

\bibitem{jain2001analysis}
R.~Jain, T.~Mullen, and R.~Hausman.
\newblock {Analysis of Paris Metro Pricing strategy for QoS with a single
  service provider}.
\newblock In {\em Proceedings of IEEE IWQoS}, pages 44--58, 2001.

\bibitem{labovitz2010internet}
C.~Labovitz, S.~Iekel-Johnson, D.~McPherson, J.~Oberheide, and F.~Jahanian.
\newblock {I}nternet inter-domain traffic.
\newblock {\em ACM SIGCOMM Computer Communication Review}, 40(4):75--86, 2010.

\bibitem{lodhi2014open}
A.~Lodhi, A.~Dhamdhere, and C.~Dovrolis.
\newblock Open peering by {I}nternet transit providers: Peer preference or peer
  pressure?
\newblock In {\em Proceedings of IEEE INFOCOM}, pages 2562--2570, 2014.

\bibitem{lodhi2015complexities}
A.~Lodhi, N.~Laoutaris, A.~Dhamdhere, and C.~Dovrolis.
\newblock Complexities in {I}nternet peering: Understanding the ``black'' in
  the ``black art''.
\newblock In {\em Proceedings of IEEE INFOCOM}, pages 1778--1786, 2015.

\bibitem{richard2014pay}
R.~T.~B. Ma.
\newblock Usage-based pricing and competition in congestible network service
  markets.
\newblock {\em IEEE/ACM Transactions on Networking}, 24(5):3084--3097, 2016.

\bibitem{ma2017pay}
R.~T.~B. Ma.
\newblock Pay or perish: The economics of premium peering.
\newblock {\em IEEE Journal on Selected Areas in Communications},
  35(2):353--366, 2017.

\bibitem{ma2013public}
R.~T.~B. Ma and V.~Misra.
\newblock The public option: A nonregulatory alternative to network neutrality.
\newblock {\em IEEE/ACM Transactions on Networking}, 21(6):1866--1879, 2013.

\bibitem{odlyzko1999paris}
A.~Odlyzko.
\newblock {Paris Metro Pricing for the {I}nternet}.
\newblock In {\em Proceedings of the 1st ACM conference on Electronic
  commerce}, pages 140--147, 1999.

\bibitem{reed2014current}
D.~P. Reed, D.~Warbritton, and D.~Sicker.
\newblock Current trends and controversies in {I}nternet peering and transit:
  Implications for the future evolution of the {I}nternet.
\newblock In {\em TPRC Conference}, 2014.

\bibitem{Rochet2003}
J.-C. Rochet and J.~Tirole.
\newblock Platform competition in two-sided markets.
\newblock {\em Journal of the European Economic Association}, 1(4):990--1029,
  2003.

\bibitem{wang2017optimal}
X.~Wang, R.~T.~B. Ma, and Y.~Xu.
\newblock On optimal two-sided pricing of congested networks.
\newblock {\em Proceedings of the ACM on Measurement and Analysis of Computing
  Systems}, 1(1):7, 2017.

\bibitem{wu2003network}
T.~Wu.
\newblock Network neutrality, broadband discrimination.
\newblock {\em Journal of Telecommunications and High Technology Law},
  2:141--179, 2003.

\bibitem{wyatt2014comcast}
E.~Wyatt and N.~Cohen.
\newblock {Comcast and Netflix reach deal on service}.
\newblock {\em The New York Times}, 23, 2014.

\end{thebibliography}
